\documentclass{aa} 
\def\scaling{1.0}
\usepackage{graphicx}
\usepackage{txfonts}
\usepackage{amsmath}
\usepackage{xcolor}

\usepackage{siunitx}
\usepackage[colorlinks=true, linkcolor=blue, urlcolor=blue, citecolor=blue]{hyperref}
\bibpunct{(}{)}{;}{a}{}{,} 
\begin{document} 

    \title{Magnetic fields in circumstellar disks}
    \subtitle{The potential of Zeeman observations}

   \author{R. Brauer\inst{1},
           S. Wolf\inst{1}, \and
           M. Flock\inst{2}
          }
          
   \authorrunning{R. Brauer et al.}

   \institute{University of Kiel, Institute of Theoretical Physics and Astrophysics,
              Leibnizstrasse 15, 24118 Kiel, Germany\\
              \email{[rbrauer;wolf]@astrophysik.uni-kiel.de}
              \and
              Jet Propulsion Laboratory, California Institute of Technology, 
              Pasadena, CA 91109, USA\\
              \email{mflock@caltech.edu}
              }


 
  \abstract
   {Recent high angular resolution polarimetric continuum observations of circumstellar disks provide new insights into their magnetic field. However, direct constraints are limited to the plane-of-sky component of the magnetic field. Observations of Zeeman split spectral lines are a potential way to enhance these insights by providing complementary information.
   }
   {We investigate which constraints for magnetic fields in circumstellar disks can be obtained from Zeeman observations of the $\SI{113}{GHz}$ CN lines. Furthermore, we analyze the conditions needed to perform these observations and their dependence on selected quantities.}
   {We simulate the Zeeman splitting with the radiative transfer (RT) code POLARIS  extended by our Zeeman splitting RT extension ZRAD, which is based on the line RT code Mol3D.
   }
   {We find that Zeeman observations of the $\SI{113}{GHz}$ CN lines provide significant insights into the magnetic fields of circumstellar disks. However, with the capabilities of recent and upcoming instruments and observatories, even spatially unresolved observations would be challenging. Nevertheless, these observations are feasible for the most massive disks with a strong magnetic field and high abundance of $CN/H$. The most restrictive quantity is the magnetic field strength, which should be at least on the order of ${\sim}\SI{1}{mG}$. In addition, the inclination of the disk should be around $60^\circ$ to preserve the ability to derive the line-of-sight (LOS) magnetic field strength and to obtain a sufficiently high circularly polarized flux. Finally, we simulate the RT of a circumbinary disk model based on a magnetohydrodynamic (MHD) simulation. We find that our analysis of the magnetic field is still applicable. However, owing to their lower circularly polarized emission, Zeeman observations of circumbinary disks with a significant separation between their stellar components ($r_\mathrm{star}\sim\SI{10}{AU}$) are more challenging than observations of circumstellar disks with a single star. }  
   {}
 

  \maketitle

\section{Introduction}\label{introduction}
The impact of magnetic fields on the formation and evolution of circumstellar disks is a matter of ongoing discussions \citep[e.g.,][]{turner_transport_2014, dudorov_fossil_2014, li_ordered_2016, khaibrakhmanov_large-scale_2017}. For instance, magnetically induced viscosity such as that caused by magnetorotational instability (MRI) depends on the degree of ionization and the structure and strength of the magnetic field inside these disks \citep{balbus_magnetohydrodynamics_2009, dzyurkevich_magnetized_2013, dudorov_fossil_2014, khaibrakhmanov_large-scale_2017}.

In recent studies, the structure and strength of magnetic fields are investigated in various environments from parsec-sized molecular clouds to the small scales of circumstellar disks \citep[e.g.,][]{pillai_magnetic_2015, li_ordered_2016}. Many of these studies rely on observations of the polarized emission of elongated dust grains that are thought to align with the longer axis perpendicular to the magnetic field lines \citep[e.g.,][]{bertrang_large-scale_2014, reissl_tracing_2014, brauer_origins_2016}. The  Chandrasekhar-Fermi method is then used to calculate the magnetic field strength in the plane-of-sky direction from the dispersions of the polarization direction and line-of-sight (LOS) velocity obtained through these observations \citep{chandrasekhar_magnetic_1953}. 

However, the observation of continuum polarization is not the only way to investigate magnetic fields in these environments. Another established method is the Zeeman splitting of spectral lines. It directly provides the magnetic field strength in the LOS direction by comparing the derivative of the net flux profile with its circularly polarized fraction \citep{crutcher_oh_1993}. Furthermore, the total magnetic field strength can be estimated using a Bayesian analysis \citep[see][]{crutcher_magnetic_2010} or by combining the Zeeman splitting results with complementary observations of aligned elongated dust grains \citep{heiles_magnetic_2012}.

A crucial challenge of Zeeman observations is the sufficient resolution of the net flux profile and its circularly polarized fraction \citep{pillai_cn_2016}. In particular, the resolution of the circularly polarized fraction is strongly affected by the strength and structure of the magnetic and velocity field and by the temperature and density distribution of the gas phase \citep{brauer_magnetic_2017}. With the sensitivity of recent instruments and observatories, only a strong circularly polarized fraction can be resolved with a sufficient signal-to-noise ratio. Therefore, successful Zeeman splitting observations are only performed on large-scale massive objects such as molecular clouds \citep[e.g.,][]{crutcher_magnetic_2012}. Nevertheless, Zeeman observations are also  expected to  provide significant insights into the structure and strength of magnetic fields in small-scale objects such as circumstellar disks. In addition, these results would be complementary to those obtained in observations of the polarized emission of elongated dust grains.

In this study, we investigate the requirements for observing the Zeeman splitting in circumstellar disks and their dependence on selected disk properties, namely the density distribution, velocity field, magnetic field strength, and inclination of the disk. To investigate the Zeeman splitting in a broad range of circumstellar disks, we consider disks around a single star and around binary stars in our simulations. For circumstellar disks around a single star, an analytical model is sufficient, whereas for circumbinary disks we use a more complex model based on magnetohydrodynamic (MHD) simulations. Based on these models, we perform radiative transfer (RT) simulations that consider the Zeeman splitting of the spectral lines of CN at $\SI{113}{GHz}$. For this purpose, we apply the RT code POLARIS \citep{reissl_radiative_2016} extended by our Zeeman splitting RT extension ZRAD \citep{brauer_magnetic_2017}, which is based on the line RT code Mol3D \citep{ober_tracing_2015}.

We structure our study as follows. We begin with a description of the RT code used for our simulations (Sect. \ref{ZRAD}). Subsequently, we introduce the Zeeman splitting formalism (Sect. \ref{zeeman_splitting}) and the characteristics of the CN lines at $\SI{113}{GHz}$ (Sect. \ref{spectral_lines}). We outline our reference disk model in Sect. \ref{model_description} and present our results in Sect. \ref{results}. Our conclusions are summarized in Sect. \ref{conclusions}.

\section{Radiative transfer}\label{ZRAD}
To perform our simulations, we apply the ZRAD extension of the 3D RT code POLARIS \citep{reissl_radiative_2016, brauer_magnetic_2017}.
This code solves the RT problem for spectral lines based on the line RT algorithm of Mol3D \citep{ober_tracing_2015} and considers the Zeeman splitting and polarization of spectral lines as well. The implementation of the Zeeman splitting is based on the works of \cite{landi_deglinnocenti_malip_1976}, \cite{schadee_zeeman_1978}, \cite{rees_stokes_1989}, and \cite{larsson_treatment_2014}. 

For each considered Zeeman split spectral line, ZRAD requires the following precalculated quantities: the energy levels and transitions taken from the Leiden Atomic and Molecular DAtabase \citep[LAMDA;][]{schoier_atomic_2005} or JPL spectral line catalog \citep{pickett_submillimeter_1998}, Land\'{e} factors of the  energy levels used, line strengths of the allowed transitions between Zeeman sublevels, and the radius of the  gas species used.

For the line shape, ZRAD includes the natural, collisional, and Doppler broadening (line shape: Voigt profile) and the magneto-optic effect \citep[line shape: Faraday-Voigt profile;][]{larsson_treatment_2014}. The Voigt and Faraday-Voigt profiles are obtained from the real and imaginary part of the Faddeeva function, respectively \citep{wells_rapid_1999}. In ZRAD, a fast and precise solution of the Faddeeva function is realized with the Faddeeva package \citep{faddeeva_package}.

\section{Zeeman splitting}\label{zeeman_splitting}
We consider the Zeeman splitting theory as outlined in Sect. 3 of our previous work \citep{brauer_magnetic_2017}. In the following, we provide a brief description of the most important physical quantities. 

By assuming that the Zeeman shift $\Delta\nu_z$ is negligible when compared to the line width $\Delta\nu$, the circularly polarized fraction $F_\text{V}$ can be written as a function of the derivative of the net flux $F_\text{I}$ as \citep[see][]{crutcher_oh_1993, brauer_magnetic_2017}
\begin{equation}
  F_\text{V}=\left(\frac{\text{d}F_\text{I}}{\text{d}\nu}\right)\Delta\nu_z\cos\theta, \label{eq:mag_field}
\end{equation}
whereas the frequency shift $\Delta\nu_z$ owing to Zeeman splitting can be calculated with
\begin{equation}
  \Delta\nu_z=\frac{B\mu_\text{B}}{h}(g^\prime M^\prime-g^{\prime\prime} M^{\prime\prime}).\label{eq:Zeeman_split}
\end{equation}
Here, $\mu_\text{B}$ is the Bohr magneton, $g$ the Land\'{e} factor, and $M$  the total angular or atomic momentum quantum number projected on the magnetic field vector $\vec{B}$. A single prime denotes the upper level; a double prime denotes the lower level of the line transition. Following Eqs. \ref{eq:mag_field} and \ref{eq:Zeeman_split}, the magnetic field strength in the LOS direction can be obtained by fitting the derivative of the net flux profile to the circularly polarized fraction of a Zeeman split spectral line.

The derived LOS magnetic field strength converges toward a reference magnetic field strength if the assumption of Eq.~(\ref{eq:mag_field}) is valid and the LOS is optically thin \citep{brauer_magnetic_2017}. The reference magnetic field strength can be calculated by averaging the LOS magnetic field strength weighted with the spectral line net flux $F_\text{I}$ along the LOS to the observer as follows:
\begin{equation}
 B_\text{LOS} = \frac{ \int_{0}^\text{obs} \left(\vec{B}(s) \cdot \vec{e_\text{LOS}}\right) F_\text{I}(s) \text{d}s }{ \int_{0}^\text{obs} F_\text{I}(s) \text{d}s } \label{eq:ref_mag_field_strength}
.\end{equation}
Here, $\vec{B}(s)$ is the magnetic field strength at a certain position, $\vec{e_\text{LOS}}$ is the unit vector in the LOS direction, and $s$ is the path length along the LOS.

\section{Spectral lines of CN at 113 GHz}\label{spectral_lines}\label{hyperfine_cn}
Zeeman observations of CN are usually performed by observing seven of the nine hyperfine transitions at ${\sim}\SI{113}{GHz}$ \citep{falgarone_cn_2008, pillai_cn_2016}. However, we limit ourselves to the spectral lines at $\SI{113.144}{GHz}$ and $\SI{113.170}{GHz}$ since they represent two extreme cases \citep[see Table \ref{tab:cn_line_qn};][]{crutcher_detection_1999, falgarone_cn_2008}. The line at $\SI{113.144}{GHz}$ has the largest Zeeman shift and one of the highest relative sensitivities to $B_\text{LOS}$ compared to the other lines. In contrast, the line at $\SI{113.170}{GHz}$ has the smallest Zeeman shift, which allows the higher LOS magnetic field strengths to be estimated  with the analysis method outlined in Sect. \ref{zeeman_splitting} \citep[see also][]{brauer_magnetic_2017}. Owing to the low Zeeman shift, the circularly polarized flux of the $\SI{113.170}{GHz}$ line requires the highest sensitivity of the seven CN lines to be detected.

The Zeeman splitting of these lines are caused by the quantized orientation of the total atomic angular momentum on the magnetic field direction. The Land\'{e} factors of the energy levels involved in these CN lines can be derived as follows \citep{bel_zeeman_1989}:
\begin{align}
  g_\text{F}&=g_\text{J}\frac{F(F+1)+J(J+1)-I(I+1)}{2F(F+1)},\\
  g_\text{J}&=\frac{J(J+1)+S(S+1)-N(N+1)}{J(J+1)}
.\end{align}
Here, $S$ is the spin quantum number ($S=0.5$) and $I$ is the atomic angular momentum quantum number ($I=1$). The quantities $N$, $J$, and $F$ are the orbital, total orbital, and total atomic angular momentum quantum numbers. In Table \ref{tab:cn_line_qn}, the values of these quantum numbers, relative intensity, and Zeeman shift per magnetic field strength are shown for each of the seven transitions.

For spectral lines caused by hyperfine transitions, the relative line strength of transitions between Zeeman sublevels can be calculated as follows \citep{larsson_treatment_2014}:
\begin{align}
  \text{If\ } \Delta F &= +1:\nonumber\\
  \Delta &M_F=0: &&S_{M^\prime,M^{\prime\prime}}=\frac{3\left((F+1)^2-M_F^2\right)}{2(F+1)(2F+1)(2F+3)},\\
  \Delta &M_F=\pm1: &&S_{M^\prime,M^{\prime\prime}}=\frac{3(F+1\pm M_F)(F+2\pm M_F)}{4(F+1)(2F+1)(2F+3)}.\\[0.5em]
  \text{If\ } \Delta F &= 0:\nonumber\\
  \Delta &M_F=0: &&S_{M^\prime,M^{\prime\prime}}=\frac{3M_F^2}{F(F+1)(2F+1)}, \label{eq:rel_line_strength_pi}\\
  \Delta &M_F=\pm1: &&S_{M^\prime,M^{\prime\prime}}=\frac{3(F\mp M_F)(F+1\pm M_F)}{4F(F+1)(2F+1)}. \label{eq:rel_line_strength_sigma}\\[0.5em]
  \text{If\ } \Delta F &= -1:\nonumber\\
  \Delta &M_F=0: &&S_{M^\prime,M^{\prime\prime}}=\frac{3(F^2-M_F^2)}{2F(2F-1)(2F+1)},\\
  \Delta &M_F=\pm1: &&S_{M^\prime,M^{\prime\prime}}=\frac{3(F\mp M_F)(F-1\mp M_F)}{4F(F-1)(2F+1)}.
\end{align}
Here, $M_F$ is the total atomic angular momentum quantum number $F$ projected on the magnetic field vector $\vec{B}$.

\begin{table*}
    \centering
    \caption{Orbital, total orbital, and total atomic angular momentum quantum numbers for the seven CN lines at $\SI{113}{GHz}$ that provide strong intensities \citep{crutcher_detection_1999, falgarone_cn_2008}. The last three columns show the relative intensity $RI$, Zeeman shift $Z$, and their multiplication for the seven CN lines as presented in  \cite{falgarone_cn_2008}.}
    \label{tab:cn_line_qn}
    \renewcommand{\arraystretch}{1.2}
    \begin{tabular}{llllllllll}
        \hline
        \hline
        $\nu_0\ [\text{GHz}]$ & $N^\prime$ & $J^\prime$ & $F^\prime$ & $N^{\prime\prime}$ & $J^{\prime\prime}$ & $F^{\prime\prime}$ & $RI$ & $Z [\SI{}{Hz/\micro G}]$ & $RI\cdot Z$ \\
        \hline
        $113.144$ & $1$ & $1/2$ & $1/2$ & $0$ & $1/2$ & $3/2$ & 8 & 2.18 & 17.4 \\
        $113.170$ & $1$ & $1/2$ & $3/2$ & $0$ & $1/2$ & $1/2$ & 8 & -0.31 & 2.5 \\
        $113.191$ & $1$ & $1/2$ & $3/2$ & $0$ & $1/2$ & $3/2$ & 10 & 0.62 & 6.2 \\
        $113.488$ & $1$ & $3/2$ & $3/2$ & $0$ & $1/2$ & $1/2$ & 10 & 2.18 & 21.8 \\
        $113.490$ & $1$ & $3/2$ & $5/2$ & $0$ & $1/2$ & $3/2$ & 27 & 0.56 & 15.1 \\
        $113.499$ & $1$ & $3/2$ & $1/2$ & $0$ & $1/2$ & $1/2$ & 8 & 0.62 & 5.0 \\
        $113.508$ & $1$ & $3/2$ & $3/2$ & $0$ & $1/2$ & $3/2$ & 8 & 1.62 & 13.0 \\
        \hline
    \end{tabular}
    \renewcommand{\arraystretch}{1}
\end{table*}

\section{Model description}\label{model_description}
\begin{table}
 \centering
 \caption{Overview of parameters for our reference circumstellar disk model.}
 \label{tab:parameter}
 \renewcommand{\arraystretch}{1.2}
 \begin{tabular}{lll}
  \hline
  \hline
  \multicolumn{3}{c}{\textit{Central star}} \\
  \hline
  Effective temperature  &  $T_\text{star}$ & $\SI{6000}{K}$ \\
  Stellar radius  &  $R_\text{star}$ & $\SI{2}{R_\odot}$ \\
  Stellar mass  &  $M_\text{star}$ & $\SI{0,7}{M_\odot}$ \\
  \hline
  \multicolumn{3}{c}{\textit{Disk model}} \\
  \hline
  Distance to star/disk  &  $d$ & $\SI{100}{pc}$ \\
  Inner radius  &  $R_\text{in}$ & $\SI{0.1}{AU}$ \\
  Outer radius  &  $R_\text{ou}$ & $\SI{300}{AU}$ \\
  Scale height  &  $h_0$ & $\SI{10}{AU}$ \\
  Characteristic radius  &  $R_\text{ref}$ & $\SI{100}{AU}$ \\
  Radial density decrease  &  $a$ & $2.625$ \\
  Disk flaring  &  $b$ & $1.125$ \\
  Cells in $r$-direction  &  $n_r$ & $100$ \\
  Step width factor in $r$  &  $\text{sf}$ & $1.05$ \\
  Cells in $\theta$-direction  &  $n_\theta$ & $91$ \\
  Cells in $\phi$-direction  &  $n_\phi$ & $180$ \\
  Inclination & $i$ & $90^\circ\ (\text{edge{-}on})$ \\
  \hline
  \multicolumn{3}{c}{\textit{Gas}} \\
  \hline
  Abundance  &  $\text{CN}/\text{H}$ & $\SI{4e-9}{}$ \\
  Spectral resolution & $\Delta \nu_\text{res}$ & $\SI{60}{kHz}\ (\SI{160}{m/s})$\\
  Turbulent velocity & $v_\text{turb}$ & $\SI{100}{m/s}$ \\
  Gas mass  &  $M_\text{gas}$ & $\SI{e-2}{M_\odot}$ \\
  Gas-to-dust mass ratio & $M_\text{gas}:M_\text{dust}$ & $100:1$ \\
  \hline
  \multicolumn{3}{c}{\textit{Dust}} \\
  \hline
  Minimum dust grain size  &  $a_\text{min}$ & $\SI{5}{nm}$ \\
  Maximum dust grain size  &  $a_\text{max}$ & $\SI{2}{\micro m}$ \\
  \hline
  \multicolumn{3}{c}{\textit{Magnetic field}} \\
  \hline
  At inner disk edge  &  $B_\text{in}(z=H)$ & $\SI{10}{G}$ \\
  At outer disk edge  &  $B_\text{ou}(z=H)$ & $\SI{1}{mG}$ \\
  \hline
 \end{tabular}
 \renewcommand{\arraystretch}{1}
\end{table}

\subsection{Disk}\label{disk}
For our reference circumstellar disk model, we consider a density distribution with a radial decrease based on the work of \citet{hayashi_structure_1981} for the minimum mass solar nebular. Combined with a vertical distribution due to hydrostatic equilibrium similar to the work of \citet{shakura_black_1973}, we obtain the following equation:
\begin{equation}
  \rho_\text{disk}=\rho_0 \left( \frac{R_\text{ref}}{\overline{\omega}} \right)^{a} \exp\left(-\frac{1}{2}\left[\frac{z}{H(\overline{\omega})}\right]^2 \right). \label{eq:disk}
\end{equation}
Here, $\overline{\omega}$ is the radial distance from the central star in the disk midplane, $z$ is the distance from the midplane of the disk, $R_\text{ref}$ is a reference radius, and $H(\overline{\omega})$ is the scale height. The density $\rho_0$ is derived from the disk (gas) mass. The scale height is a function of $\overline{\omega}$ as follows:
\begin{equation}
  H(\overline{\omega})=h_0 \left(\frac{\overline{\omega}}{R_\text{ref}}\right)^{b}. \label{eq:disk2}
\end{equation}
The parameters $a$ and $b$ set the radial density profile and the disk flaring, respectively. The extent of the disk is constrained by the inner radius $R_\text{in}$ and outer radius $R_\text{ou}$. For our circumstellar disk model, we consider a gas mass of $M_\text{gas}=\SI{e-2}{M_\odot}$, which is a typical value for Class II YSO disks \citep{andrews_circumstellar_2005, robitaille_interpreting_2007}. Furthermore, we assume that the observer is $\SI{100}{pc}$  from the disk, which corresponds to a few circumstellar disks that can be found inside of $\SI{100}{pc}$ (CQ Tau, \citealt{miroshnichenko_dust_1999}; PDS 66, \citealt{silverstone_formation_2006}; Hen 3-600A, \citealt{andrews_truncated_2010}; TW Hya, \citealt{andrews_ringed_2016}). However, more circumstellar disks can be considered by decreasing the simulated flux in the figures of Sect.~\ref{results}, which is equivalent to increasing the distance. Nevertheless, this will not change the influence of physical properties on the observable quantities of the disk.
 
\subsection{Gas}
For the simulation of a spectral line, we assume local thermodynamic equilibrium (LTE). In particular, we assume that the gas and dust are in thermal equilibrium, and the dust is in thermal equilibrium with the surrounding radiation field. Non-LTE approximations such as the one based on the local velocity gradient (LVG) may be more accurate for circumstellar disks. However, additional information about collision rates would be needed. In addition, as shown for the $\mathrm{HCO}^+(4-3)$ transition in the work of \citet{ober_tracing_2015}, differences between the LTE and non-LTE approximations should at most amount to a factor of 2. Therefore, we expect that these differences are negligible compared to the uncertainties of physical conditions in circumstellar disks. We assume an abundance of CN to hydrogen which is in agreement with observations \citep[see Table \ref{tab:parameter};][]{falgarone_cn_2008}. Furthermore, we use velocity channels with a spectral resolution that is achievable with state-of-the-art instruments/observatories such as the Atacama Large Millimeter/submillimeter Array  \citep[ALMA; see Table \ref{tab:parameter};][]{alma_technical_handbook}.

\subsection{Dust}\label{dust}
The RT code POLARIS allows one to consider  the dust continuum and the molecular line radiation. However, as we simulate the spectral lines of CN at $\SI{113}{GHz}$ (corresponding to ${\sim}\SI{2650}{\micro m}$), the thermal emission of the dust is negligible in comparison to the spectral line emission. Nevertheless, we need to define the optical properties of the dust grains to calculate the disk temperature profile. In particular, we assume compact, homogeneous, and spherical dust grains, consisting of 62.5\% silicate and 37.5\% graphite (MRN dust, \citealt{mathis_size_1977}; optical properties from \citealt{weingartner_dust_2001}). For the size distribution of the dust grains we apply
\begin{equation}
  \text{d}n(a)\propto a^{-3.5} \text{d}a, \quad a_\text{min} < a < a_\text{max}, \label{eqn:dust}
\end{equation}
where $n(a)$ is the number of dust particles with a specific dust grain radius $a$. We assume a minimum and maximum dust grain size that is typical for ISM grains \citep[see Table \ref{tab:parameter};][]{mathis_size_1977}. Significant grain growth and dust settling are not considered since their influence on the temperature distribution should be negligible compared to the uncertainties of other disk parameters. The total dust mass of the disk model is considered to be $\SI{0.01}{M_\mathrm{gas}}$.

\subsection{Magnetic field}\label{magnetic_field}
\begin{figure}
  \centering
 \resizebox{\scaling\hsize}{!}{\includegraphics{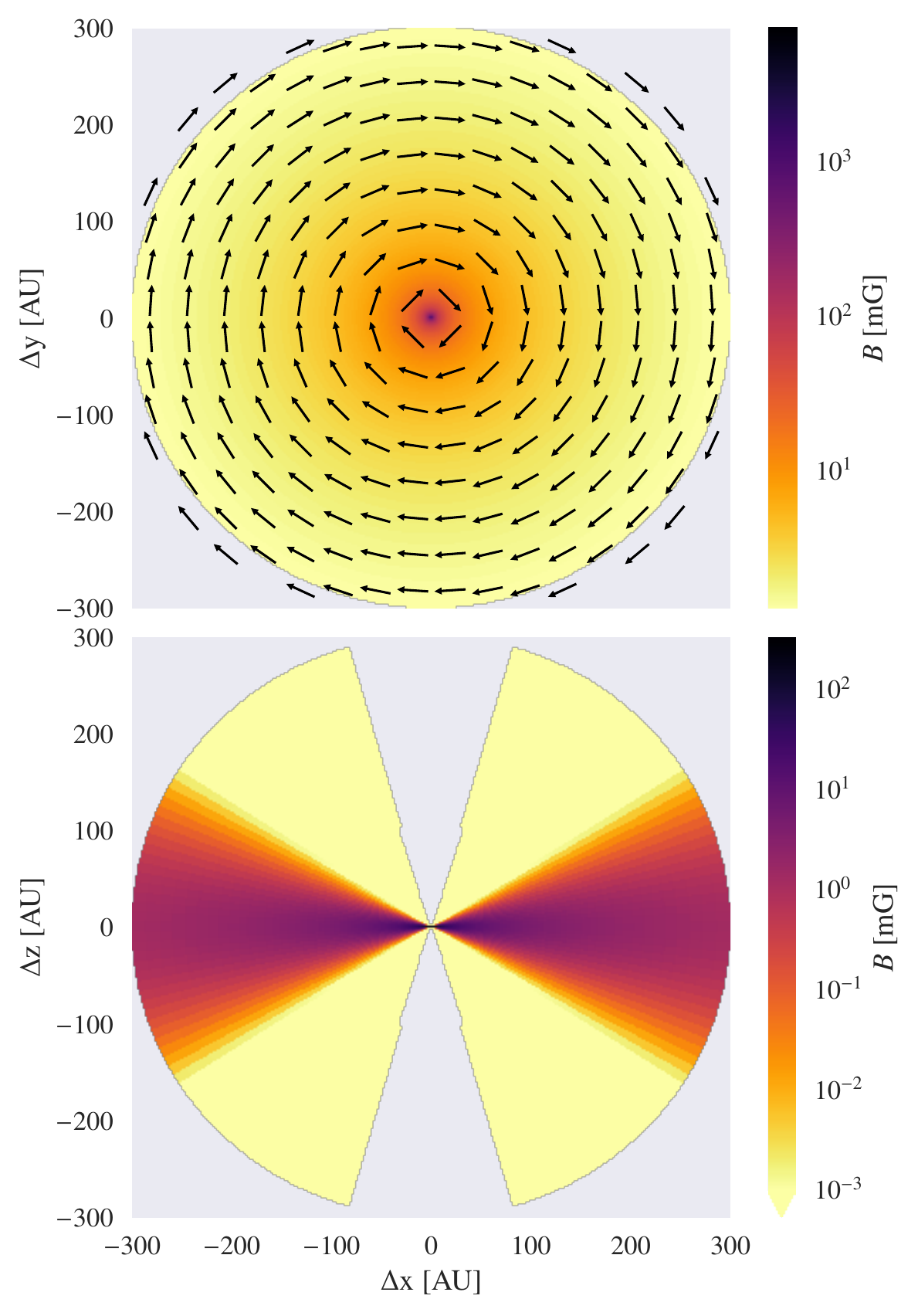}}
  \caption{Magnetic field strength in the disk midplane (\textit{upper}) and as a vertical cut through the disk (\textit{lower}). The arrows indicate the direction but not the strength of the magnetic field.}
  \label{fig:magnetic_field_cuts}
\end{figure}

We assume a toroidal magnetic field structure for our circumstellar disk model, which is in agreement with observations and MHD simulations of young circumstellar disks \citep{flock_turbulence_2012, segura-cox_magnetic_2015, bertrang_magnetic_2017}. Our magnetic field strength distribution depends on the density and reproduces the same radial behavior as shown in Fig. 4 of the work from \cite{dudorov_fossil_2014} as follows:
\begin{align}
 \vec{B}(\rho) &= B_0\left(\frac{\rho_\text{disk}}{\rho_0}\right)^{\alpha}\vec{e}_\varphi, \label{eq:magnetic_field} \\
 \alpha &=  0.44.
\end{align}
Here, $B_0$ and $\rho_0$ are normalization factors set to define the magnetic field strength at the inner and the outer edges of the circumstellar disk model, $\vec{e}_\varphi$ is the unit vector in azimuthal direction, and $\rho_\text{disk}$ is the density as defined in Eq.~(\ref{eq:disk}). To be consistent with the work by \cite{dudorov_fossil_2014}, we constrain the magnetic field strength distribution with the field strengths at the inner and outer disk radii at $z=H$ (see Table \ref{tab:parameter} and Eq.~(\ref{eq:disk2})). Figure \ref{fig:magnetic_field_cuts} shows the magnetic field strength in the disk midplane (upper) and as a vertical cut through the disk (lower). The exponent of $0.44$ is somewhat different to the value obtained in other studies for molecular clouds \citep[$\alpha\sim2/3$;][]{mestel_magnetic_1966, crutcher_magnetic_2010}. However, for circumstellar disks, a value of $\alpha$ that differs significantly from $0.44$ results in a magnetic field strength at the disk edges that is not consistent with either the magnetic field of the central star or the surrounding interstellar medium.


\subsection{Velocity field}\label{velocity_field}
We consider Keplerian rotation for the velocity field in our circumstellar disk model. Thus, the velocity at a certain radial distance from the central star in the disk midplane $\overline{\omega}$ can be written as follows:
\begin{align}
 \vec{v} = \sqrt{\frac{GM_\text{star}}{\overline{\omega}}}\vec{e}_\varphi.
\end{align}
Here, $G$ is the gravitational constant, $M_\text{star}$ is the mass of the central star ($M_\text{star}=\SI{0,7}{M_\odot}$). To take gas motion on scales smaller than the spatial resolution into account, we consider a turbulent velocity of $v_\text{turb}=\SI{100}{m/s}$, which is in agreement with observations \citep{pietu_probing_2007, chapillon_chemistry_2012}.

\subsection{Heating source}\label{heating_source}
The primary heating source is a central pre-main sequence star. The star is characterized by its effective temperature $T_\text{star}=\SI{6000}{K}$ and radius $R_\text{star}=\SI{2}{R_\odot}$.

\subsection{Circumbinary disk}\label{mhd_simulation}
In addition to  the reference model of a circumstellar disk, we also perform simulations for a circumbinary disk model that is based on a snapshot from a global 3D non-ideal MHD stratified simulation. The numerical and physical setup is based on the circumstellar disk model as described in our previous work (see model \texttt{D2G\_e-2} in Sect. 2 and Fig. 7 in \citealt{flock_gaps_2015}). For the current setup, we replace the single star potential with a time variable potential from a binary source. Both stars rotate around the center of mass at a distance of $\SI{5}{AU}$ with a constant rotation frequency of $\Omega=\sqrt{G M_{c}/(\SI{5}{AU})}$, where $M_c=\SI{0.5}{M_\odot} + \SI{0.5}{M_\odot}$ (for additional parameters see Table \ref{tab:parameter_mhd}). The separation in the $\phi$ plane amounts to $180^\circ$. We first let the system relax to a new equilibrium condition. The binary potential is constantly perturbing the disk. However, we observe that the magnetic field stabilizes this effect. For the circumbinary disk model, we use the snapshot that is taken after a time evolution of 150 inner orbits.

\begin{figure*}[htpb]
 \centering
 \resizebox{\hsize}{!}{\includegraphics[page=1]{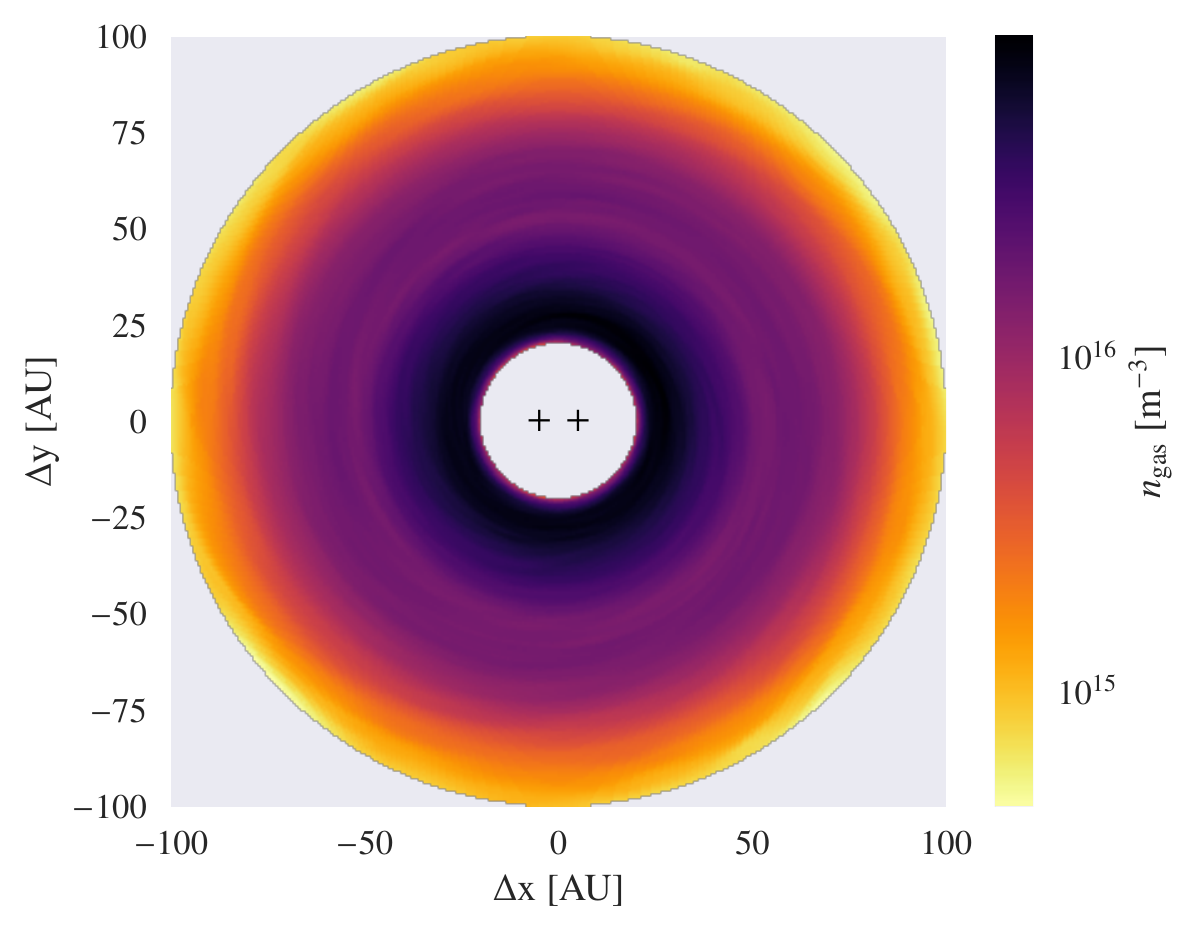} \quad
                       \includegraphics[page=2]{figures/MHD_midplane_cuts.pdf}}
 \resizebox{\hsize}{!}{\includegraphics[page=4]{figures/MHD_midplane_cuts.pdf} \quad
                       \includegraphics[page=3]{figures/MHD_midplane_cuts.pdf}}
  \caption{Gas number density (\textit{upper left}), magnetic field strength (\textit{upper right}), and velocity field (\textit{lower left}) in the disk midplane as well as the magnetic field strength shown as a vertical cut through the disk (\textit{lower right}). The arrows indicate the direction but not the strength of the corresponding quantity. The black $+$ signs mark the positions of the binary stars.}
  \label{fig:magnetic_field_cuts_mhd}
\end{figure*}

\begin{table}
 \centering
 \caption{Overview of parameters for our circumbinary disk model based on a MHD simulation. For grid parameters see \cite{flock_gaps_2015}.}
 \label{tab:parameter_mhd}
 \renewcommand{\arraystretch}{1.2}
 \begin{tabular}{lll}
  \hline
  \hline
  \multicolumn{3}{c}{\textit{Binary stars (identical)}} \\
  \hline
  Effective temperature  &  $T_\text{star}$ & $\SI{4500}{K}$ \\
  Stellar radius  &  $R_\text{star}$ & $\SI{0.41}{R_\odot}$ \\
  Stellar mass  &  $M_\text{star}$ & $\SI{0.5}{M_\odot}$ \\
  Distance to disk center & $r_\text{star}$ & $\SI{5}{AU}$ \\
  \hline
  \multicolumn{3}{c}{\textit{Disk model}} \\
  \hline
  Distance to star/disk  &  $d$ & $\SI{100}{pc}$ \\
  Inner radius  &  $R_\text{in}$ & $\SI{20}{AU}$ \\
  Outer radius  &  $R_\text{ou}$ & $\SI{100}{AU}$ \\
  Inclination & $i$ & $90^\circ\ (\text{edge{-}on})$ \\
  \hline
  \multicolumn{3}{c}{\textit{Gas}} \\
  \hline
  Abundance  &  $\text{CN}/\text{H}$ & $\SI{4e-9}{}$ \\
  Spectral resolution & $\Delta \nu_\text{res}$ & $\SI{60}{kHz}\ (\SI{160}{m/s})$\\
  Turbulent velocity & $v_\text{turb}$ & $\SI{100}{m/s}$ \\
  Gas mass  &  $M_\text{gas}$ & $\SI{2.5e-2}{M_\odot}$ \\
  Gas-to-dust mass ratio & $M_\text{gas}:M_\text{dust}$ & $100:1$ \\
  \hline
  \multicolumn{3}{c}{\textit{Dust}} \\
  \hline
  Minimum dust grain size  &  $a_\text{min}$ & $\SI{5e-3}{\micro m}$ \\
  Maximum dust grain size  &  $a_\text{max}$ & $\SI{2}{\micro m}$ \\
  \hline
 \end{tabular}
 \renewcommand{\arraystretch}{1}
\end{table}

\section{Results}\label{results}
In this section, we present the results and analysis of our simulations. At first, we investigate whether observations of Zeeman split CN lines at $\SI{113}{GHz}$ can be used to estimate the structure and strength of the magnetic field in circumstellar disks (see Sects. \ref{result_ideal_observations} and \ref{result_real_observations}). Subsequently, we analyze the impact of selected disk properties on the requirements of these observations (see Sects. \ref{result_mag_field} to \ref{result_kepler}). This is followed by a study of the observability of regions with significantly altered magnetic field strengths such as dead zones (see Sect. \ref{result_dead-zone}). Then, we discuss the potential of Zeeman observations of other spectral lines in circumstellar disks (see Sect. \ref{result_other_spectral_lines}). Finally, we replace our circumstellar disk model with a circumbinary disk model and consider a MHD simulation instead of an analytical approach to investigate the potential of Zeeman observations in a more realistic environment (see Sect. \ref{result_zeeman_binary}).

\subsection{Ideal observations}\label{result_ideal_observations}
\begin{figure*}
  \def\fscaling{1.0}
  \centering
  \resizebox{\fscaling\hsize}{!}{\includegraphics{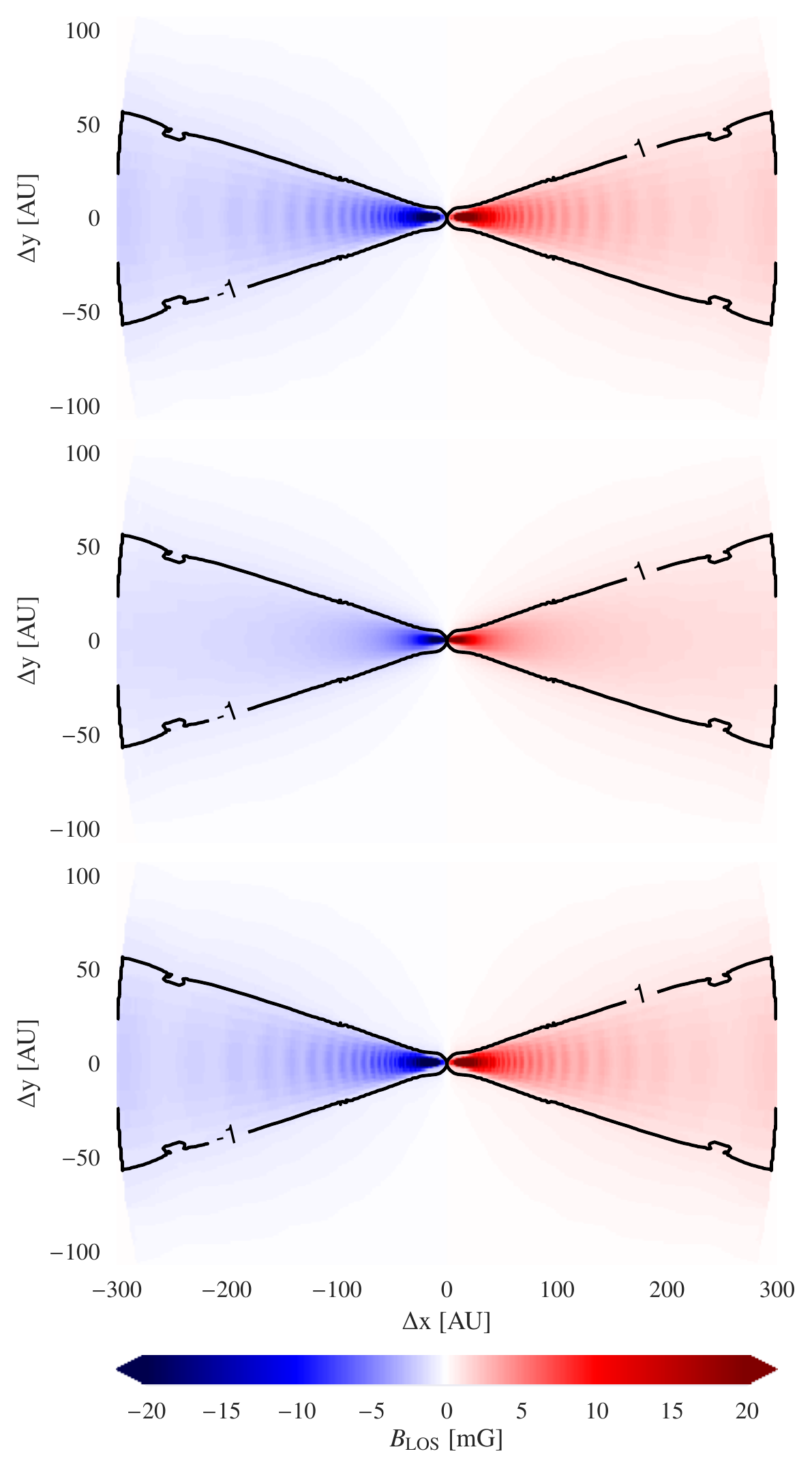}\qquad
                        \includegraphics{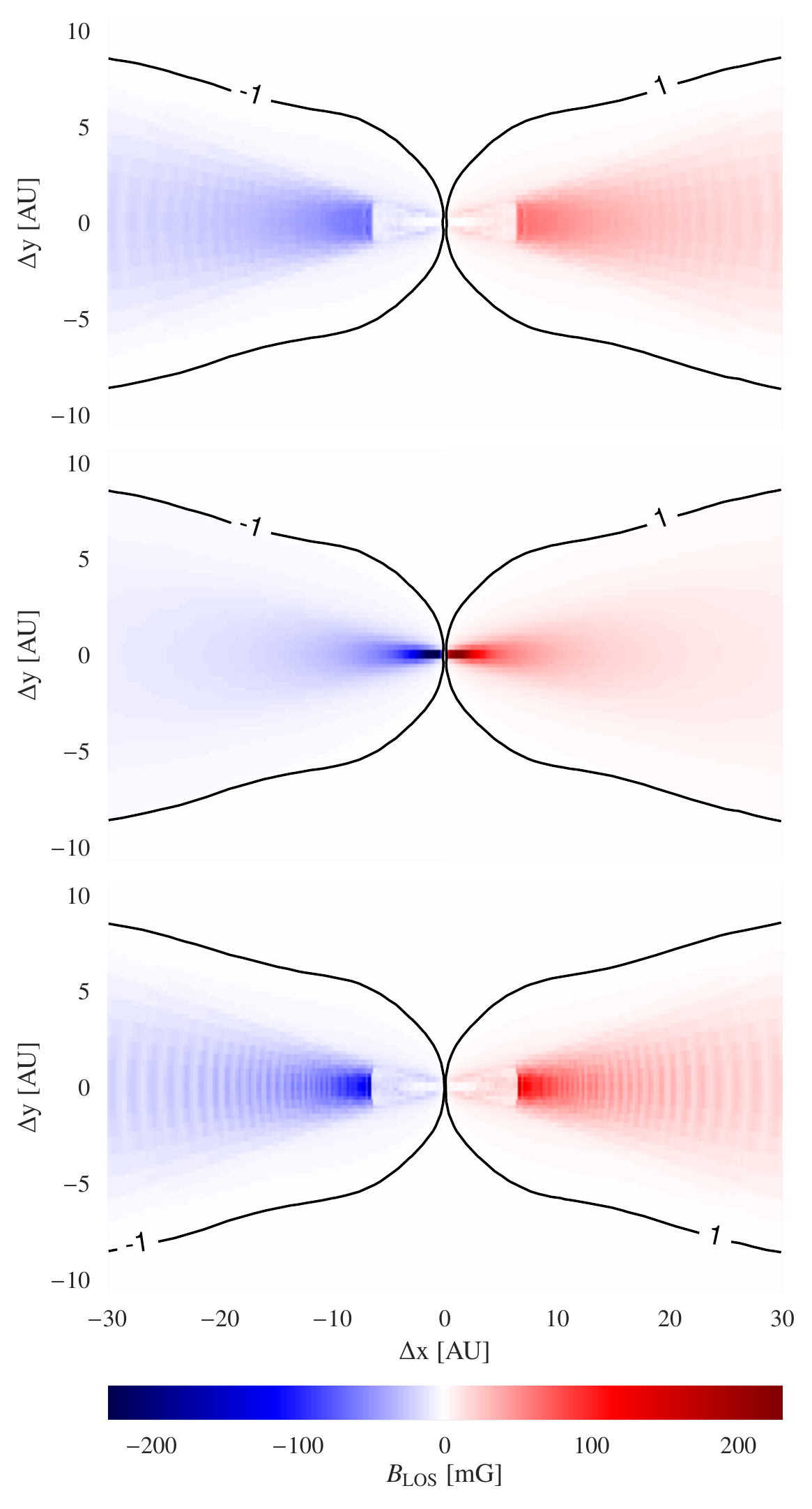}}
  \caption{Derived (\textit{upper} and \textit{lower}) and reference (\textit{middle}) LOS magnetic field strength calculated for each pixel of the simulated velocity channel map of the reference circumstellar disk model ($\SI{113.144}{GHz}$ (\textit{upper}) and $\SI{113.170}{GHz}$ (\textit{lower}) CN line emission). The images on the \textit{right} side are zoomed in on the center by a factor of 10. The faint wave-like structure in the \textit{upper} and \textit{lower} images comes from the continuous analytical model and the calculation of the derivative of the spectral line profile, which is composed of a finite number of velocity channels.}
  \label{fig:cn_best_resolution}
\end{figure*}

\begin{figure}
  \centering
  \resizebox{\scaling\hsize}{!}{\includegraphics{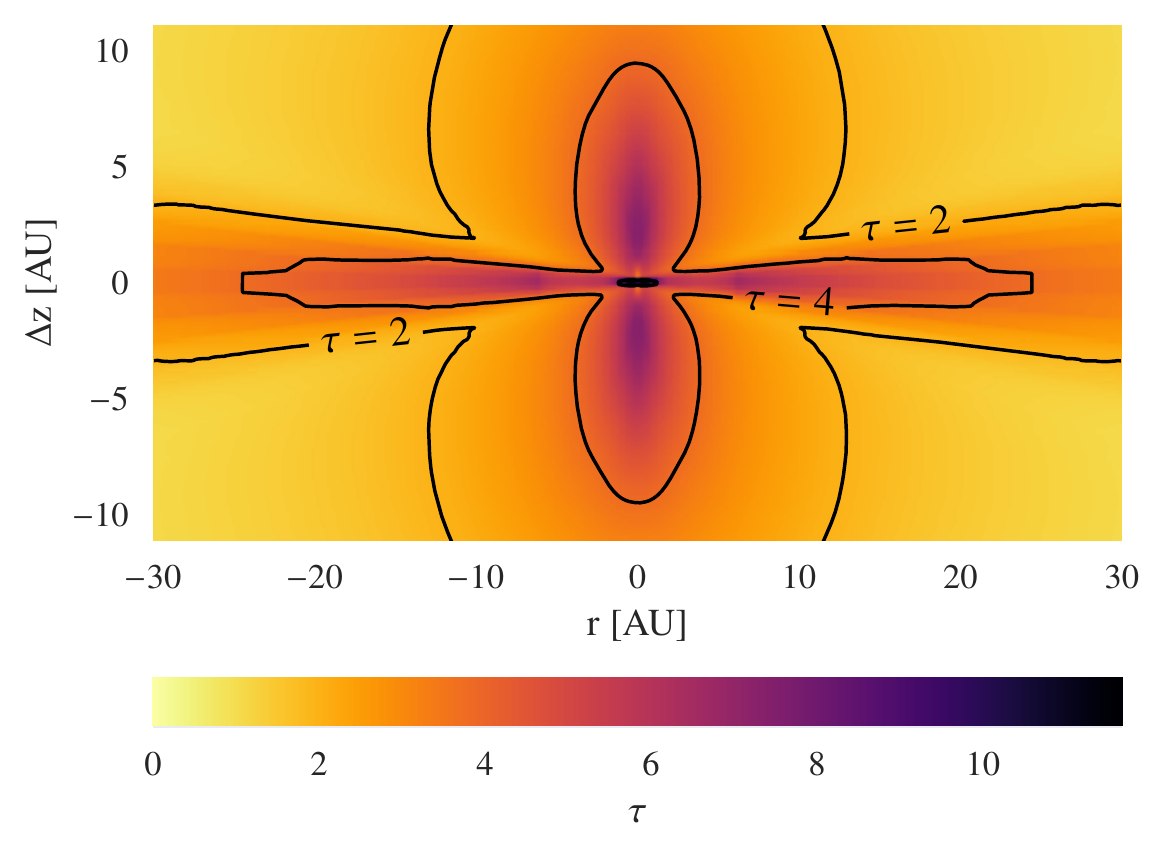}}
  \caption{Optical depth of the $\SI{113.144}{GHz}$ CN line emission of the reference circumstellar disk model. The optical depth is calculated along the LOS through the whole disk by averaging the optical depth over the velocity channels weighted with the corresponding net flux. The high optical depth above and below the midplane does not only correspond to high gas densities, but is mainly caused by the low dispersion of the velocity along the LOS. The contour lines correspond to optical depths of $2$, $4$, and $8$.}
  \label{fig:cn_best_resolution_tau}
\end{figure}

We begin our study with the analysis of ideal observations, i.e., we neglect any constraints resulting from the limited spatial resolution and sensitivity of real observing instruments/observatories. We simulate the emission of the $\SI{113.144}{GHz}$ and $\SI{113.170}{GHz}$ CN lines from the circumstellar disk model obtained in Sect. \ref{disk}. As illustrated in Fig.~\ref{fig:cn_best_resolution} (upper and lower), the LOS magnetic field strengths derived from the circularly polarized fraction of the two CN lines clearly show the toroidal structure of the magnetic field. In addition, the LOS magnetic field strengths are similar to the field strengths taken directly from our disk model (see Eq.~(\ref{eq:ref_mag_field_strength}) and Fig.~\ref{fig:cn_best_resolution}, middle). However, the magnetic field strength in the innermost region close to the star cannot be reproduced. This is caused by the very high magnetic field strength and optical depth in this region (see Figs. \ref{fig:magnetic_field_cuts} and \ref{fig:cn_best_resolution_tau}). In contrast to the $\SI{113.144}{GHz}$ CN line, the derived LOS magnetic field strength from the $\SI{113.170}{GHz}$ CN line reaches greater field strengths, which is expected from its smaller Zeeman shift \citep[see Table \ref{tab:cn_line_qn};][]{brauer_magnetic_2017}:
\begin{align}
    \SI{113.144}{GHz}& \text{\ CN\ line:}\nonumber\\
    &\frac{\nu_z}{B}=\SI{2.18}{\frac{Hz}{\micro G}} \quad\Rightarrow\quad  B_\text{LOS,max}\sim\SI{70}{mG}\\
    \SI{113.170}{GHz}& \text{\ CN\ line:}\nonumber\\
    &\frac{\nu_z}{B}=\SI{0.31}{\frac{Hz}{\micro G}} \quad\Rightarrow\quad  B_\text{LOS,max}\sim\SI{160}{mG}
\end{align}
However, the increase in the maximum LOS magnetic field strength is smaller than the decrease in the Zeeman shift. This can be explained by the fact that the derived LOS magnetic field strength is averaged along the LOS, taking also regions with magnetic field strength lower than the maximum value into account.

In summary, we find that CN Zeeman observations of typical circumstellar disks have the potential to identify a possible toroidal structure of the magnetic field and correctly estimate the averaged LOS magnetic field strength of most regions in the disk. However, such an observation would need a spatial resolution on the order of ${\sim}\SI{1}{AU}$ and a sensitivity per resolution element on the order of ${\sim}\SI{0.01}{\micro Jy}$. However, these requirements are far beyond the capabilities of cutting-edge instruments/observatories such as ALMA \citep[see][]{alma_technical_handbook}.

\begin{figure}
    \centering
    \resizebox{\scaling\hsize}{!}{\includegraphics{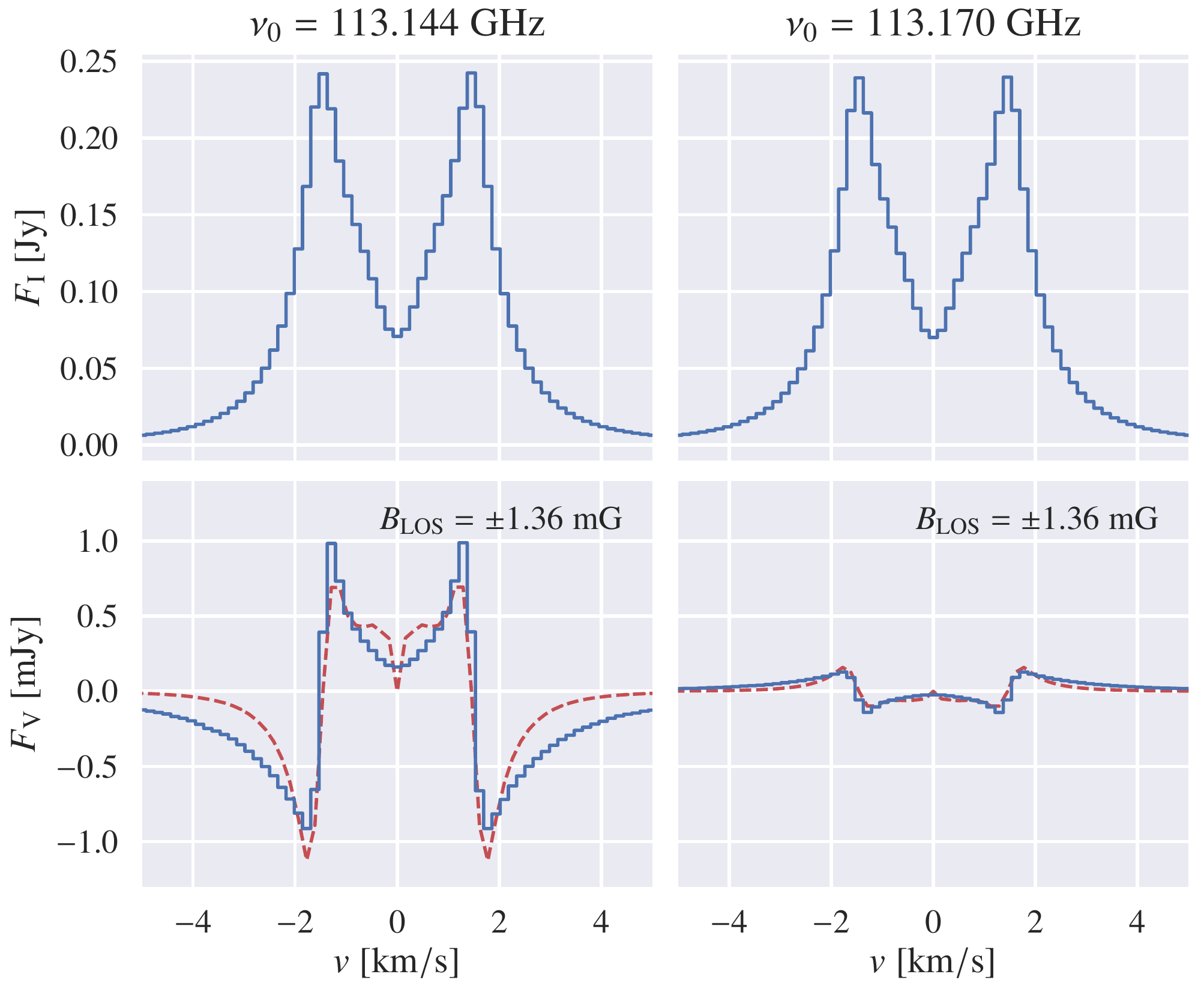}}
    \caption{Net flux ($F_\text{I}$, \textit{upper}) and circularly polarized fraction ($F_\text{V}$, \textit{lower}) simulated for the reference circumstellar disk model. In the \textit{lower} images, the derivative of $F_\text{I}$ is shown and fitted to $F_\text{V}$ to estimate the LOS magnetic field strength (red dashed line). The profiles on the \textit{left} and \textit{right} side are from the $\SI{113.144}{GHz}$ and $\SI{113.170}{GHz}$ CN line, respectively.}
    \label{fig:cn_profile}
\end{figure}

If the sensitivity is high enough to spectrally resolve the circularly polarized flux of the whole disk, the LOS magnetic field strength can also be estimated from spatially unresolved spectral line profiles (see Fig.~\ref{fig:cn_profile}). For reasons of symmetry, an ideal toroidal magnetic field structure would result in a zero net LOS magnetic field strength. However, we take advantage of the Keplerian rotation in the disk. The spectral line emission from one side of the disk with a velocity towards the observer is shifted to higher frequencies, whereas the emission from the other side with a velocity away from the observer is shifted to lower frequencies (see Fig.~\ref{fig:cn_profile}). Since velocity and magnetic field both have  a toroidal structure, the emission observed at frequencies lower or higher than the rest line peak are almost entirely related to regions in the disk with a magnetic field strength directed towards or away from the observer. Therefore, we can estimate an average LOS magnetic field strength for both sides of the disk independently. This can also be used to observe asymmetries between the magnetic field structure of both sides. Nevertheless, as expected from the spatial averaging, the resulting field strengths are lower than the values obtained from spatially resolved images (see Figs. \ref{fig:cn_best_resolution} and \ref{fig:cn_profile}). However, they still allow  constraints on the magnetic field strength in circumstellar disks to be derived. 

The difference between the estimated LOS magnetic field strengths of the two CN lines is negligible (see Figs. \ref{fig:cn_best_resolution} and \ref{fig:cn_profile}). However, the circularly polarized flux of the $\SI{113.170}{GHz}$ line is almost an order of magnitude lower than the flux of the $\SI{113.144}{GHz}$ line (see Fig.~\ref{fig:cn_profile}, lower). As the aim of this study is to find the best conditions under which  to perform Zeeman observations, we focus our further investigations on the $\SI{113.144}{GHz}$ CN line. 

\subsection{Simulation of real observations}\label{result_real_observations}
In the next step, we compare the CN line emission of the reference circumstellar disk model with typical sensitivities of ALMA obtained with the corresponding observing tool \citep{alma_ot, alma_technical_handbook}. With a spectral resolution of $\SI{60}{kHz}$ (see Table \ref{tab:parameter}), we obtain a sensitivity of $\Delta F \approx \SI{10}{mJy}$ for a full polarization observation of the $\SI{113.144}{GHz}$ CN line with a total observation time of about three hours. Because ALMA is not yet capable of measuring circularly polarized flux, this sensitivity is calculated under the assumption that it is the same for observations of circular polarization as for linear polarization. However, the observing mode for the circularly polarized flux is planned for the future \citep{alma_proposers_guide}. According to Fig.~\ref{fig:cn_profile}, the sensitivity of ALMA is only sufficient to resolve the net flux, but not the circularly polarized fraction. Therefore, our next aim is to investigate selected disk properties and observational conditions at which future CN Zeeman observations of circumstellar disks are most likely to succeed. We consider disk properties that are least constrained and/or have been found to vary most from object to object. These are the strength and structure of the magnetic field, abundance of CN compared to hydrogen, inclination of the circumstellar disk, and mass of the central star, i.e., the strength of the Keplerian velocity field.

\subsection{Magnetic field strength}\label{result_mag_field}
\begin{figure}
    \centering
    \resizebox{\scaling\hsize}{!}{\includegraphics{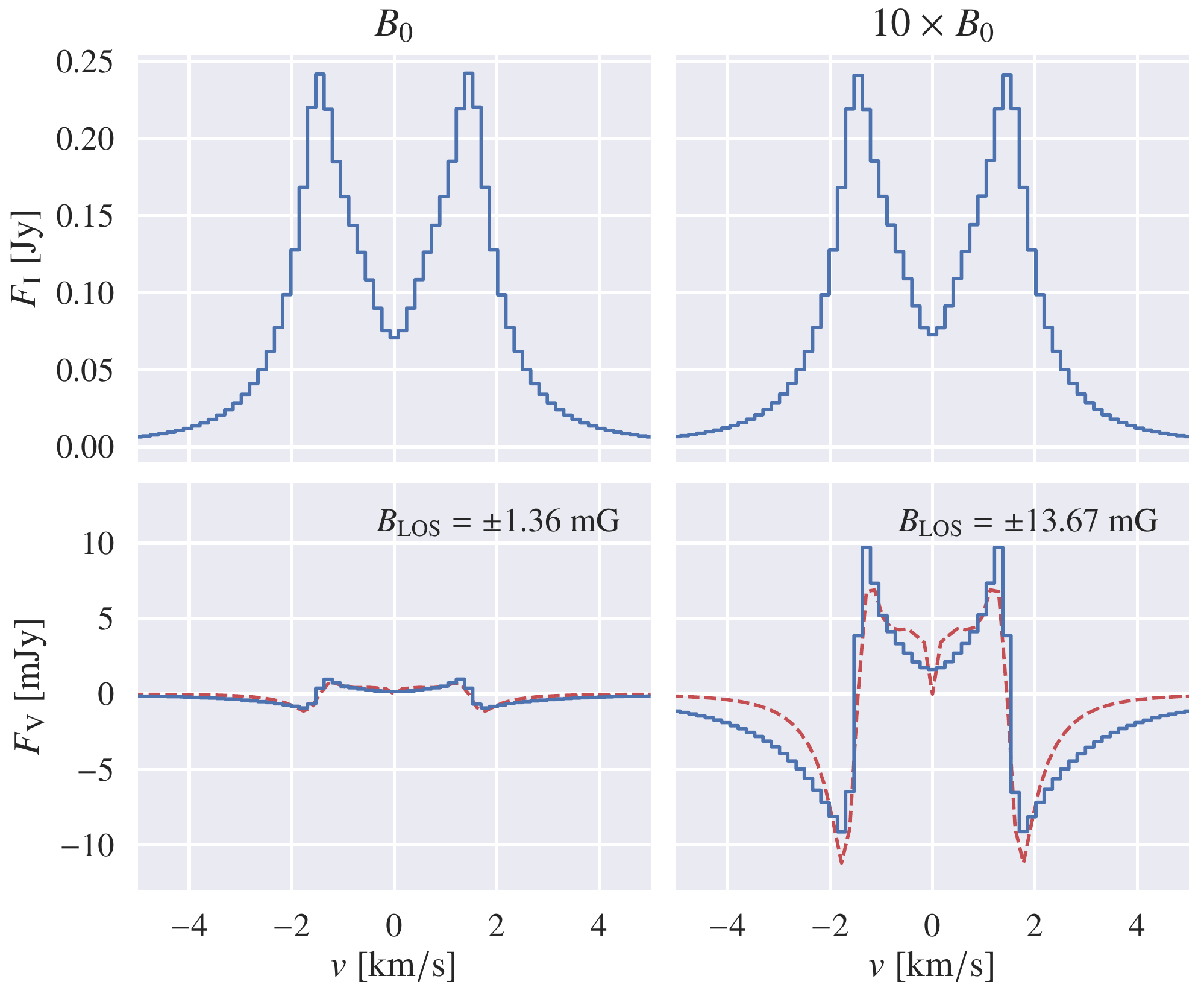}}
    \caption{Net flux ($F_\text{I}$, \textit{upper}) and circularly polarized fraction ($F_\text{V}$, \textit{lower}) simulated for the reference circumstellar disk model ($\SI{113.144}{GHz}$ CN line emission). In the \textit{lower} images, the derivative of $F_\text{I}$ is shown and fitted to $F_\text{V}$ to estimate the LOS magnetic field strength (red dashed line). The simulated profiles on the \textit{right} side use the reference disk model with an increased magnetic field strength. Throughout the disk, the magnetic field strength is 10 times higher than assumed in Eq.~(\ref{eq:magnetic_field}) (Sect. \ref{magnetic_field}).}
    \label{fig:cn_profile_mag_1e1}
\end{figure}

An increase in the magnetic field strength increases the Zeeman splitting of the spectral line, i.e., the circularly polarized flux, hence the  sensitivity required to perform successful Zeeman observations decreases. For a quantitative analysis of the impact of an increased magnetic field strength, we investigate the influence of a higher magnetic field strength by performing the same simulations as shown in Fig.~\ref{fig:cn_profile} with a 10 times higher magnetic field strength throughout the disk model. By comparing the left and right columns in Fig.~\ref{fig:cn_profile_mag_1e1}, it can be seen that the circularly polarized flux increases by almost the same amount as the magnetic field. Nevertheless, the increase in the magnetic field strength is still not high enough to have a significant impact on the net flux profile. Since the sensitivity needed to detect the circularly polarized fraction is the limiting factor in Zeeman observations, circumstellar disks with high magnetic field strengths should be preferred over those with low field strengths. 

\subsection{Abundance}\label{result_abundance}
\begin{figure*}
    \centering
    \resizebox{\hsize}{!}{\includegraphics{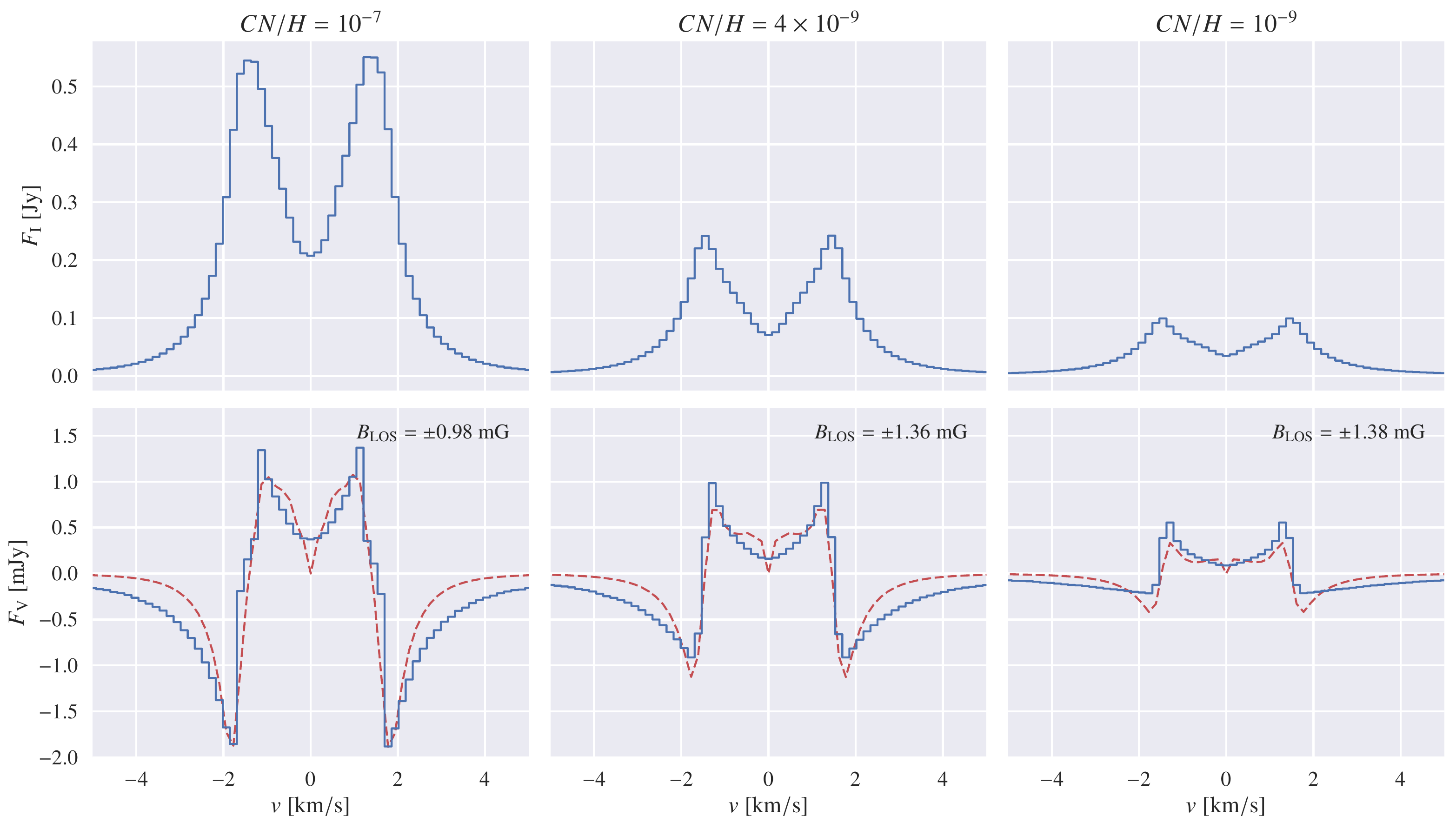}}
    \caption{Net flux ($F_\text{I}$, \textit{upper}) and circularly polarized fraction ($F_\text{V}$, \textit{lower}) simulated for the reference circumstellar disk model ($\SI{113.144}{GHz}$ CN line emission). In the \textit{lower} images, the derivative of $F_\text{I}$ is shown and fitted to $F_\text{V}$ to estimate the LOS magnetic field strength (red dashed line). The profiles on the \textit{left}, \textit{middle}, and \textit{right} side are simulated with an abundance of $CN/H=\SI{e-7}{}$, $CN/H=\SI{4e-9}{}$, and $CN/H=\SI{e-9}{}$, respectively.}
    \label{fig:cn_profile_abundance_1e-7}
\end{figure*}

An increase in the $CN/H$ abundance directly increases the CN line emission of a circumstellar disk (optically thin case). However, not every LOS in our circumstellar disk model is optically thin in the $\SI{113}{GHz}$ CN lines (see Fig.~\ref{fig:cn_best_resolution_tau}). Therefore, an increase in the abundance will result in a less than linear increase in the spectral line emission of the observing instrument/observatory. To investigate how the abundance influences the line emission, we performed simulations of the reference disk model as performed for Fig.~\ref{fig:cn_profile} and used either a 25 times higher or 4 times lower abundance of CN to hydrogen (see Fig.~\ref{fig:cn_profile_abundance_1e-7}). By comparing the spectral line profiles for different abundances, it can be seen that the relative change in the net flux is significantly smaller than the variation in the abundance. Therefore, our circumstellar disk model is optically thick for abundances higher  than $CN/H\sim\SI{e-9}{}$. As a consequence, with increasing abundance, regions of high magnetic field strength become hidden and do not contribute to the resulting spectral line profile. Thus, the relative change in the circularly polarized fraction is even smaller than the variation in the net flux and the estimated magnetic field strength decreases with increasing abundance. 

In summary, the abundance of CN to hydrogen only has  a weak impact on the likelihood of  performing a successful Zeeman observation of the $\SI{113}{GHz}$ CN lines emitted from circumstellar disks. Also, a variation in the circumstellar disk mass has the same influence as a change in the $CN/H$ abundance and is therefore not analyzed separately.

\subsection{Inclination}\label{result_inclination}
\begin{figure*}
    \centering
    \resizebox{\hsize}{!}{\includegraphics{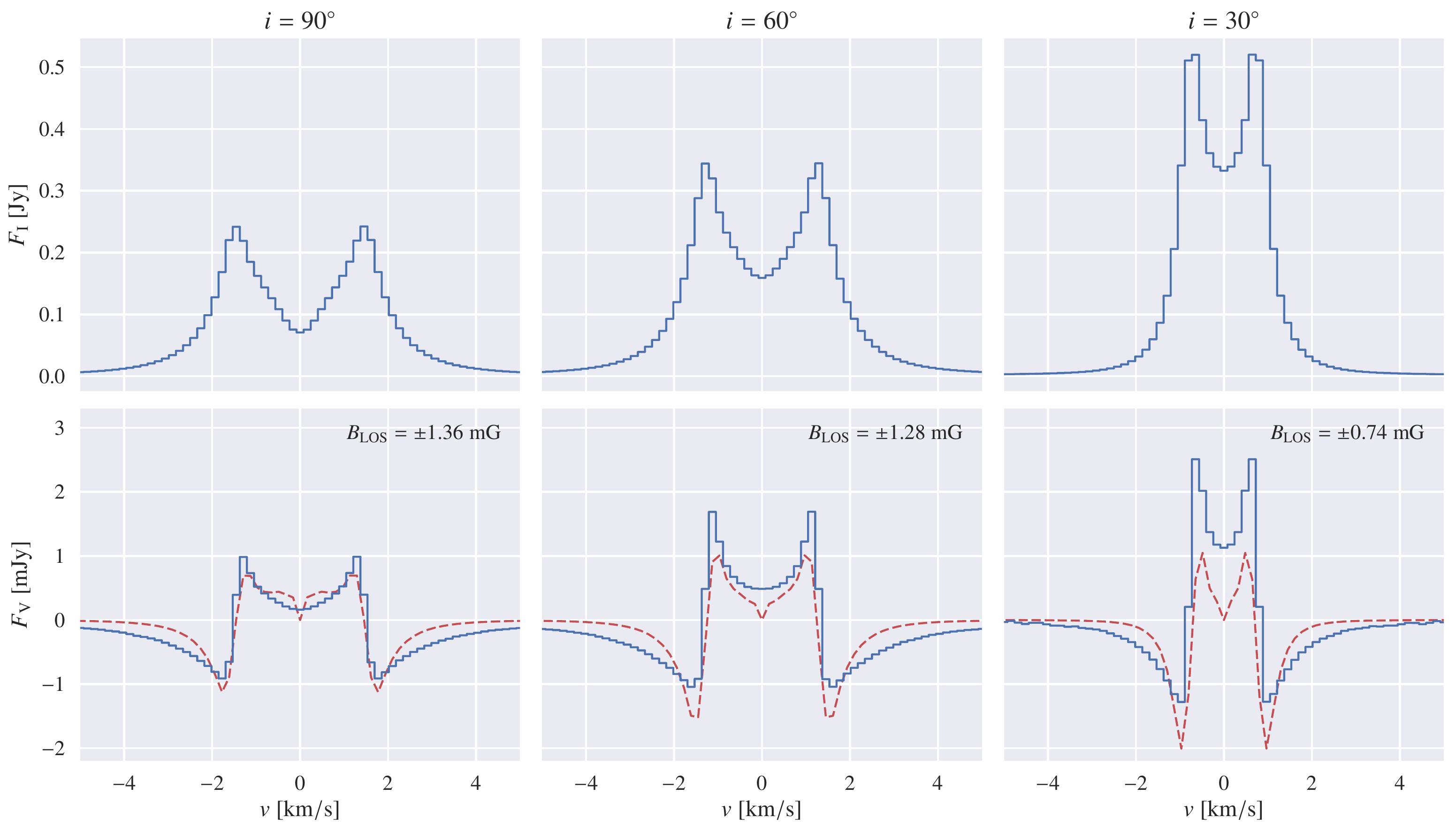}}
    \caption{Net flux ($F_\text{I}$, \textit{upper}) and circularly polarized fraction ($F_\text{V}$, \textit{lower}) simulated for the reference circumstellar disk model ($\SI{113.144}{GHz}$ CN line emission). In the \textit{lower} images, the derivative of $F_\text{I}$ is shown and fitted to $F_\text{V}$ to estimate the LOS magnetic field strength (red dashed line). The profiles on the \textit{left}, \textit{middle}, and \textit{right} side are simulated with an inclination of $i=90^\circ$, $i=60^\circ$, and $i=30^\circ$, respectively.}
    \label{fig:cn_profile_inc}
\end{figure*}

A variation in the circumstellar disk inclination has various impacts on Zeeman observations. The magnetic field strength in the LOS direction, velocity field in the LOS direction, optical depth, and emitting cross section of the disk vary with the inclination as a sine or cosine function (optically thin case). However, as shown before, our circumstellar disk model is not optically thin. Therefore, we investigate how the inclination influences the derived LOS magnetic field strength and the required sensitivity to resolve the circularly polarized fraction. More specifically, we perform simulations of the reference disk model with an inclination from $5^\circ$ up to $90^\circ$ in steps of $5^\circ$. The resulting spectra for $90^\circ$, $60^\circ$, and $30^\circ$ are illustrated in Fig.~\ref{fig:cn_profile_inc}. The upper images illustrate that the net flux profiles increase with decreasing inclination as expected from the larger emitting cross  section of the disk. This effect is strong enough to cause an increase in the circularly polarized fraction even if the LOS component of the magnetic field decreases with lower inclinations.

\begin{figure}
    \centering
    \resizebox{\scaling\hsize}{!}{\includegraphics{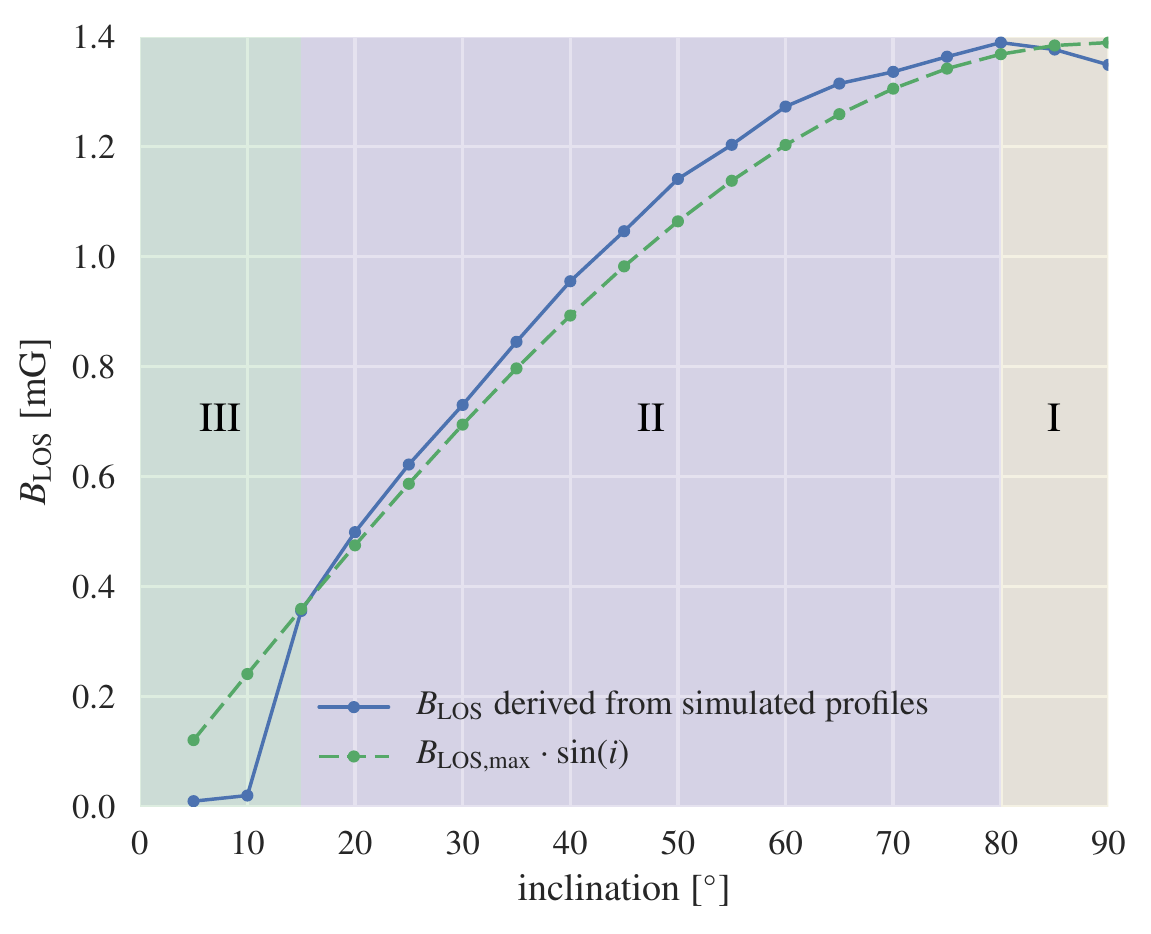}}
    \caption{Derived LOS magnetic field strength calculated from one half of the simulated $F_\text{I}$ and $F_\text{V}$ spectra of the reference circumstellar disk model with different inclinations ($\SI{113.144}{GHz}$ CN line emission). For comparison, a sine function is plotted to show the ideal behavior of the magnetic field component in the LOS direction with inclination. The three roman numerals highlight regions whose LOS magnetic field strengths show different behavior.}
    \label{fig:cn_profile_inc_study}
\end{figure}

Because of the smaller optical depth at lower inclinations, the contribution of regions with higher magnetic field strengths to the average LOS magnetic field strength derived from the circularly polarized emission is stronger. Consequently, the LOS magnetic field strength should vary with inclination as a modified sine function. To investigate this, Fig.~\ref{fig:cn_profile_inc_study} illustrates the average LOS magnetic field strength from our simulations and a sine function fitted to these results. In Fig.~\ref{fig:cn_profile_inc_study}, we can distinguish between three regions whose average LOS magnetic field strength  show  different behavior. In  regions I and II, the lower inclination and therefore lower optical depth allow parts of the disk with high magnetic field strength to contribute to the average LOS magnetic field strength. In region I, this even outshines the decrease in the magnetic field strength in the LOS direction with decreasing inclination. However, the deviation from the ideal sine function is rather small in these two regions. As a result, if the magnetic field in a circumstellar disk has a mainly toroidal structure, the inclination can be safely used to constrain the maximum average LOS magnetic field strength. In region III, the Kepler shifted spectral line peaks are too close to each other to allow a sufficiently accurate estimation of the LOS magnetic field strength. Therefore, we propose that Zeeman observations should not be used on disks with an inclination smaller than ${\sim}15^\circ$. Since real circumstellar disks are more complex than our analytical model, this inclination limit could be even higher.

In summary, depending on the specific inclination of the disk, Zeeman observations of the $\SI{113}{GHz}$ CN lines provide different advantages and disadvantages. Disks seen close to edge-on are best suited to reconstructing the total magnetic field strength and the structure of the magnetic field including asymmetries. In contrast, disks with lower inclinations provide a higher circularly polarized flux and are therefore more likely to be observed.

\subsection{Kepler rotation}\label{result_kepler}
\begin{figure*}
    \centering
    \resizebox{\hsize}{!}{\includegraphics{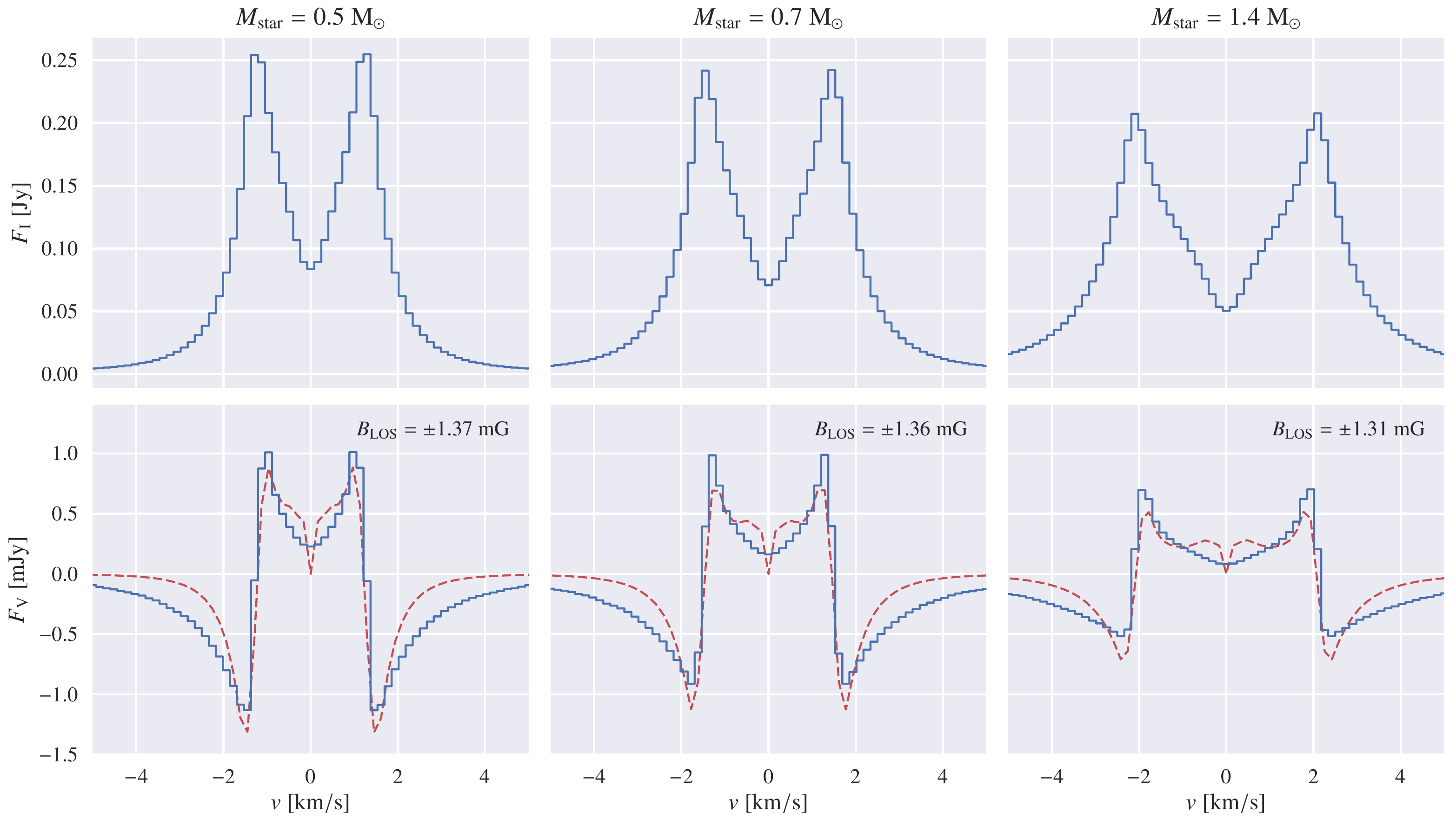}}
    \caption{Net flux ($F_\text{I}$, \textit{upper}) and circularly polarized fraction ($F_\text{V}$, \textit{lower}) simulated for the reference circumstellar disk model ($\SI{113.144}{GHz}$ CN line emission). In the \textit{lower} images, the derivative of $F_\text{I}$ is shown and fitted to $F_\text{V}$ to estimate the LOS magnetic field strength (red dashed line). The profiles on the \textit{left}, \textit{middle}, and \textit{right} side are simulated with a mass of the central star of $M_\text{star}=\SI{0.5}{M_\odot}$, $M_\text{star}=\SI{0.7}{M_\odot}$, and $M_\text{star}=\SI{1.4}{M_\odot}$, respectively.}
    \label{fig:cn_profile_kepler}
\end{figure*}

In our reference disk model, we assumed a certain mass of the central star to calculate the Keplerian velocity field in the disk. The velocity shift is a main factor in the derivation of the LOS magnetic field strength from spectral line profiles. Therefore, we investigate how a change in the velocity field influences the potential for observing the Zeeman splitting of the $\SI{113}{GHz}$ CN lines emitted from circumstellar disks. As illustrated in Fig.~\ref{fig:cn_profile_kepler}, a significant change in the mass of the central star and therefore of the velocity field has only a minor impact on the derived LOS magnetic field strength and the circularly polarized emission. If we combine this finding with the inclination dependence of the reconstructed magnetic field, we conclude that a smaller (larger) central star mass increases (decreases) the minimum inclinations that is required for successful Zeeman observation of circumstellar disks (see Sect. \ref{result_inclination}).

\subsection{Dead zone}\label{result_dead-zone}
In Sect.~\ref{result_inclination}, we found that in the case of small inclinations, the derived average LOS magnetic field strength is also influenced by the high magnetic field strengths that exist in dense regions of the disk. In addition, various studies predict that the gas is weakly ionized in these dense regions, since the column density is high enough to shield it from ionizing radiation such as cosmic rays \citep{umebayashi_fluxes_1981, turner_dead_2008, dudorov_fossil_2014}. Such a region is called a dead zone and is expected to be located at a distance  between $\SI{1}{AU}$ and $\SI{20}{AU}$ from the disk with a vertical extent of up to ${\sim}\SI{0.5}{AU}$ \citep{dzyurkevich_magnetized_2013, dudorov_fossil_2014}. Therefore, we investigate whether a small region with a low magnetic field strength close to the star can be detected with Zeeman observations of the $\SI{113}{GHz}$ CN lines. We perform simulations with our reference disk, but mimic the dead zone  by modifying the magnetic field strength as follows:
\begin{equation}
  \vec{B}=
  \begin{cases}
    \hspace{0.1cm} \left(0,\,0,\,0\right) \quad&\text{if}\ \overline{\omega} \leq \SI{5}{AU}\ \text{and}\ z \leq \SI{0.5}{AU}\\
    \hspace{0.1cm} \vec{B(\rho)} \quad&\text{if}\ \overline{\omega} > \SI{5}{AU}\ \text{or}\ z > \SI{0.5}{AU}\\
  \end{cases}
  \label{eq:mag_field_dead-zone}
.\end{equation}
Here, $\vec{B(\rho)}$ is the magnetic field strength distribution defined in Eq.~(\ref{eq:magnetic_field}) (Sect. \ref{magnetic_field}). Real dead zones should have a non-zero magnetic field strength. However, we investigate the most extreme case with our approach.

\begin{figure}
    \centering
    \resizebox{\scaling\hsize}{!}{\includegraphics{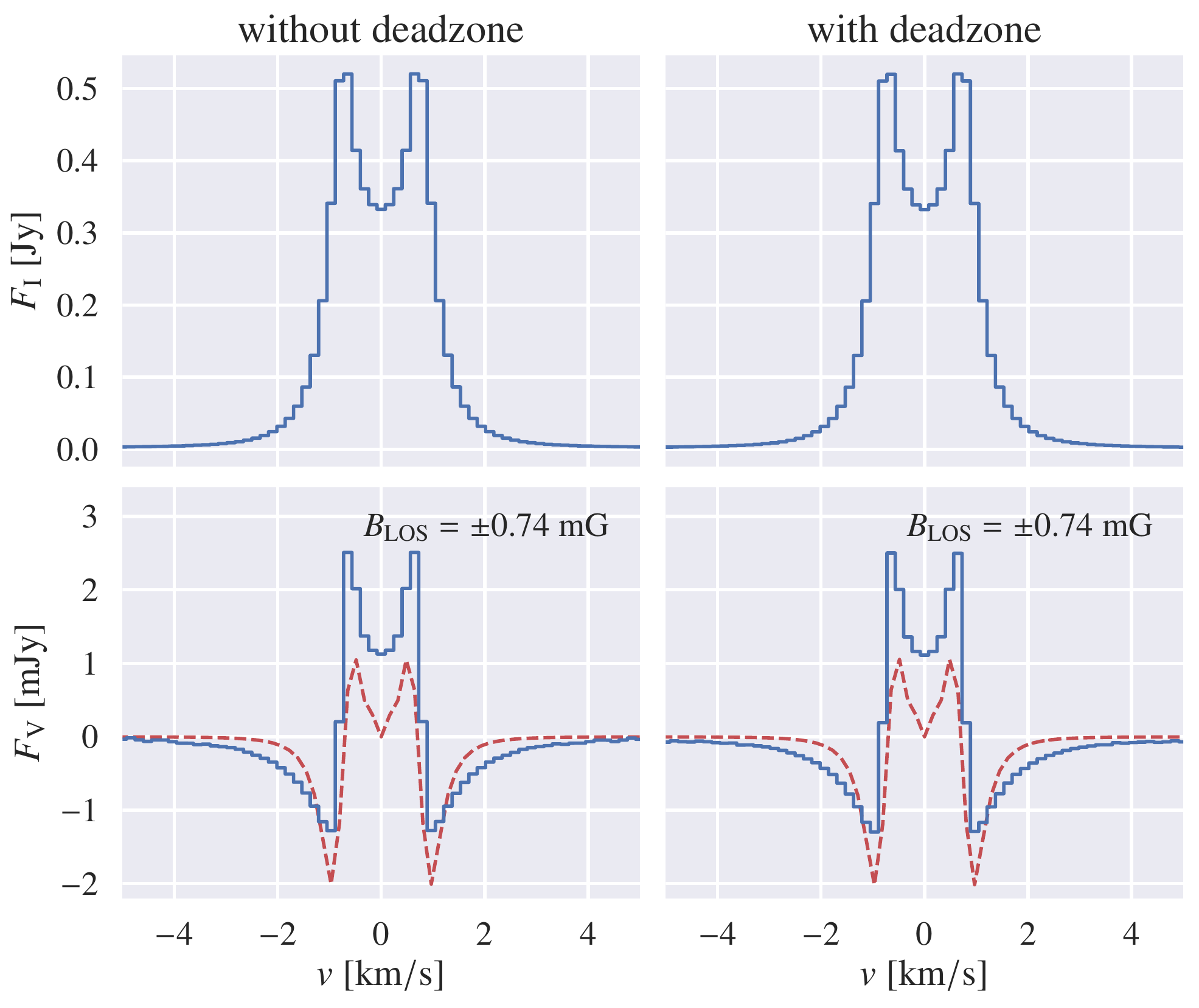}}
    \caption{Net flux ($F_\text{I}$, \textit{upper}) and circularly polarized fraction ($F_\text{V}$, \textit{lower}) simulated for the reference circumstellar disk model ($\SI{113.144}{GHz}$ CN line emission). In the \textit{lower} images, the derivative of $F_\text{I}$ is shown and fitted to $F_\text{V}$ to estimate the LOS magnetic field strength (red dashed line). The profiles on the \textit{left} are simulated with the reference disk model with an inclination of $30^\circ$. The simulated profiles on the \textit{right} side use the reference disk model with an inclination of $30^\circ$ and a dead zone close to the star. The extent of this dead zone ranges from the inner disk edge  to a radial distance of ${\sim}\SI{5}{AU}$ and from the midplane  to a vertical distance of ${\sim}\SI{0.5}{AU}$ in both directions.}
    \label{fig:cn_dead-zone_spectrum}
\end{figure}

As illustrated in Fig.~\ref{fig:cn_dead-zone_spectrum}, the dead zone has only a negligible influence on the net flux and circularly polarized emission. Therefore, the derived LOS magnetic field strength remains the same. An influence of the dead zone can only be seen in Zeeman observations with a very high spatial resolution. Hence, spatially unresolved Zeeman observations are not suitable for detecting significant variations in the magnetic field strength if they are limited to small regions close to the star. In addition, our reference circumstellar disk model has a fairly low gas mass compared to the masses used by several studies that investigate dead zones \citep[e.g.,][]{dudorov_fossil_2014}. In disks with higher gas masses, the dead zones would be hidden even more efficiently, which further reduces the chance of a detection.

\subsection{Other spectral lines}\label{result_other_spectral_lines}
Zeeman observations of the $\SI{113}{GHz}$ CN lines are not the only possibility for investigating the magnetic field in the ISM. For molecular clouds, observations of the $\SI{1665}{MHz}$ and $\SI{1667}{MHz}$ OH lines and $\SI{21}{cm}$ HI line have also been performed successfully \citep{crutcher_magnetic_1999, troland_magnetic_2008, crutcher_magnetic_2010}. However, these spectral lines are not well suited to investigating the magnetic field in circumstellar disks for various reasons. At first, OH and HI are only good tracers of the cold less dense ISM, but not of the hot dense gas in a circumstellar disk \citep{crutcher_magnetic_2010}. Therefore, the emerging flux from a circumstellar disk at these spectral lines will be far too low to be detected \citep[typical $\SI{1665}{MHz}$ OH disk emission: $F_\text{I}\sim\SI{1}{\micro Jy}$; SKA sensitivity: $\Delta F_\text{I}\sim\SI{100}{\micro Jy}$;][]{ska_booklet}. Furthermore, the magnetic field that can be traced by these spectral lines will be related to the cold outer regions of the disk. Information of the magnetic field in the hot inner regions cannot be obtained. Another reason is the spatial resolution of recent radio telescopes. At the frequencies of theses spectral lines, the spatial resolution is too low to ensure that only the emission of the disk is detected \citep[e.g.,][]{crutcher_testing_2009, ska_booklet}. 

Almost any Zeeman detection in molecular clouds was made with OH, CN, or HI. Nevertheless, other species such as $\mathrm{C_2H}$, $\mathrm{SO}$, $\mathrm{C_2S}$, $\mathrm{C_4H}$, and $\mathrm{CH}$ may also be promising candidates for investigating magnetic fields \citep{crutcher_magnetic_2012}. However, these species are not  in the current study.

\subsection{Circumbinary disk}\label{result_zeeman_binary}
Most circumstellar disk models consider a single star as their sole source of radiation and gravitational potential. However, a significant fraction of stars are expected to form in multiple systems, preferentially as binaries \citep[e.g.,][]{duquennoy_multiplicity_1991, kraus_coevality_2009, wurster_impact_2017}. For example, V4046 Sgr hosts a circumbinary disk with a narrow ring at $r\sim\SI{37}{AU}$ that contains the majority of the dust mass \citep{rosenfeld_structure_2013}. For the system HH 30, an inner disk radius of ${\sim}\SI{37}{AU}$ gives a hint of the existence of a binary \citep{guilloteau_resolving_2008}. In contrast, the circumstellar disk around the GG Tau quadruple stellar system is an example of a complex environment with at least four stellar components \citep{mccabe_nicmos_2002}.

Although our analysis is applicable in the case of binary systems consisting of two individual circumstellar disks, it remains an open question whether it has to be modified in the case of circumbinary disks with a small but non-negligible separation between their stellar components ($r_\mathrm{star}\sim\SI{10}{AU}$). For this reason, we perform RT simulations for such a circumbinary disk and obtain the model through MHD simulations to take the increased complexity in the case of a circumbinary disk into account. The simulations are performed in similar way to the analytical circumstellar disk model (parameters shown in Table \ref{tab:parameter_mhd}). From the results, we derive the main differences between Zeeman observations of circumstellar disks and circumbinary disks. Furthermore, we investigate whether our previous analysis of the magnetic field strength can be applied to circumbinary disks as well.

\begin{figure}
  \centering
  \resizebox{\scaling\hsize}{!}{\includegraphics{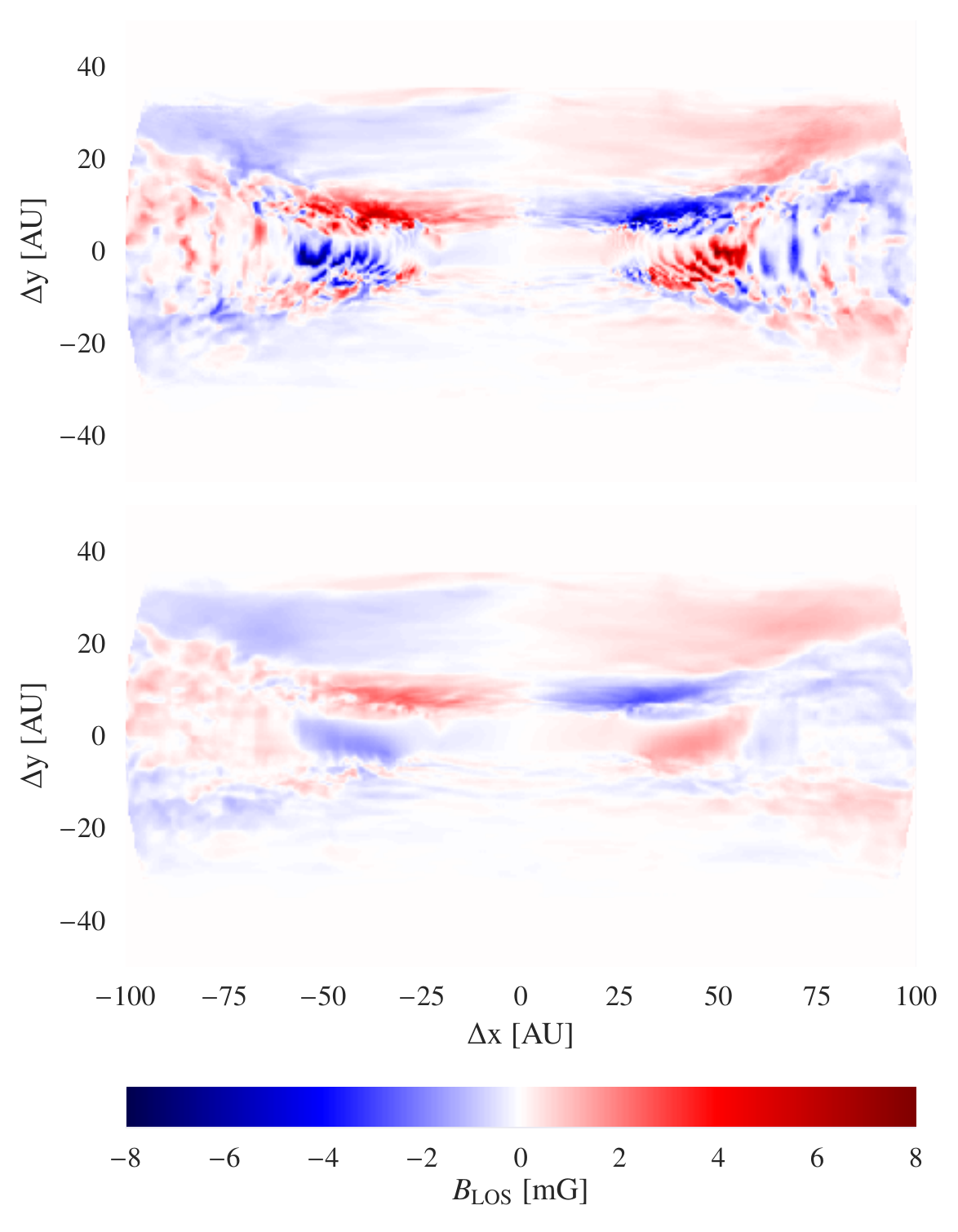}}
  \caption{Derived (\textit{upper}) and reference (\textit{lower}) LOS magnetic field strength calculated for each pixel of the simulated velocity channel map of the circumbinary disk model based on the MHD simulation described in Sect. \ref{result_zeeman_binary}.}
  \label{fig:cn_best_resolution_mhd_simulation}
\end{figure}

As illustrated in Fig.~\ref{fig:cn_best_resolution_mhd_simulation}, the estimated magnetic field strength in the LOS direction shows a toroidal structure that has different directions of circulation depending on the region of the disk. Furthermore, the estimated LOS magnetic field structure (Fig.~\ref{fig:cn_best_resolution_mhd_simulation}, upper) agrees well with the magnetic field taken out of the MHD simulation (Fig.~\ref{fig:cn_best_resolution_mhd_simulation}, lower). The increase in the derived LOS magnetic field strength compared to the field strength of the MHD simulation can be explained by the limited spectral resolution and the interaction between a spatially varying velocity and magnetic field \citep[see][]{brauer_magnetic_2017}. However, as mentioned in Sect. \ref{result_ideal_observations}, the requirements for observing a circumstellar disk with such a high spatial resolution are still beyond the capabilities of recent instruments/observatories. As with our analytical circumstellar disk model, a key to this problem may be spatially unresolved Zeeman observations. 

\begin{figure}
    \centering
    \resizebox{\scaling\hsize}{!}{\includegraphics{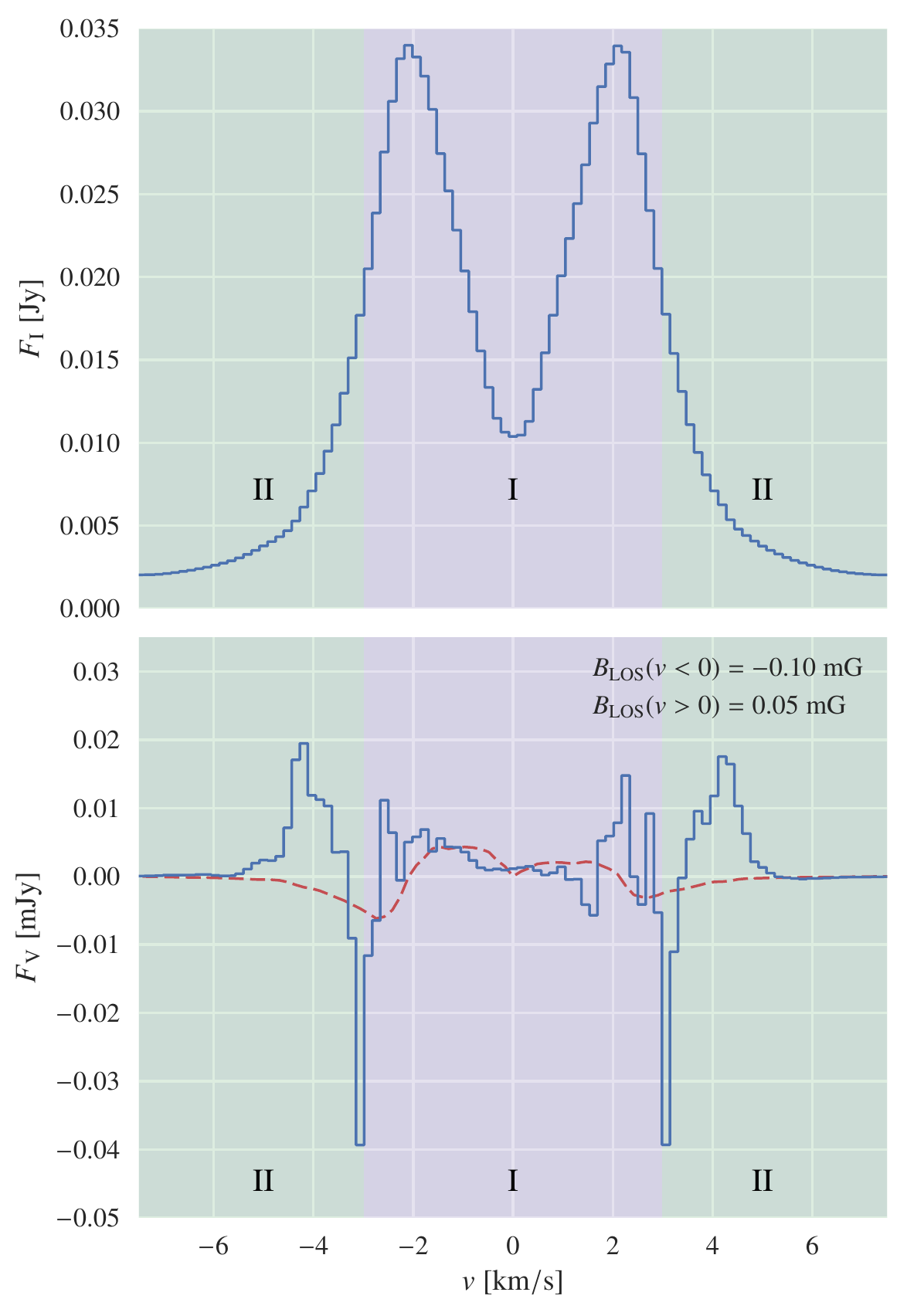}}
    \caption{Net flux ($F_\text{I}$, \textit{upper}) and circularly polarized fraction ($F_\text{V}$, \textit{lower}) simulated for the circumbinary disk model based on a MHD simulation. In the \textit{lower} image, the derivative of $F_\text{I}$ is shown and fitted to $F_\text{V}$ to estimate the LOS magnetic field strength (red dashed line). The fitting of the derivative of $F_\text{I}$ to $F_\text{V}$ is only performed for the velocity channels between $\SI{-3.2}{km/s}$ and $\SI{3.2}{km/s}$.}
    \label{fig:cn_profile_mhd_simulation}
\end{figure}

With spatially unresolved Zeeman observations, only the net flux of the circumbinary disk can be analyzed. As illustrated in Fig.~\ref{fig:cn_profile_mhd_simulation} (upper), the net flux shows the characteristic double peak structure that is expected from the Keplerian motion of the gas (similar to Fig.~\ref{fig:cn_profile} of Sect. \ref{result_ideal_observations}). However, the circularly polarized fraction is far more complex (see Fig.~\ref{fig:cn_profile_mhd_simulation}, lower). On each side (positive and negative velocities), two regions of different origin can be distinguished (see regions I and II in Fig.~\ref{fig:cn_profile_mhd_simulation}). Region I is related to the main emission of the disk that is caused by the slow-moving outer part with relatively low magnetic field strength (see Fig.~\ref{fig:magnetic_field_cuts_mhd} in Sect. \ref{magnetic_field}). A Zeeman signal (alternation between negative and positive peaks) can barely be identified in this region, due to the strong variation in the magnetic field direction combined with the line shift owing to the velocity field (see Fig.~\ref{fig:magnetic_field_cuts_mhd} in Sect. \ref{magnetic_field}, upper right). In contrast to region I, the spectral line emission of region II is caused by the fast-moving inner part of the disk with significantly higher magnetic field strength. As illustrated in Fig.~\ref{fig:magnetic_field_cuts_mhd} (upper right) in Sect. \ref{magnetic_field}, the magnetic field direction is more stable in the inner part of the disk than in the outer part. Therefore, the Zeeman signal of region II is not only stronger but also more pronounced than the signal of region I.


An estimation of the LOS magnetic field strength from the derivative of $F_\text{I}$ is only reliable for region I since it is related to the main emission of the disk (see Fig.~\ref{fig:cn_profile_mhd_simulation}). For region II, the corresponding net flux has to be extracted from the $F_\text{I}$ profile which is difficult and prone to errors. However, in real Zeeman observations of circumbinary disks, distinguishing between regions I and II for the fitting process may not be possible. Then, the derived LOS magnetic field strength differs significantly from the field strengths in different regions of the disk. In addition, the possibility of distinguishing different regions of the spectral line profile decreases with decreasing inclination, since both peaks of the net flux are getting closer to each other. 

In contrast to our analytical disk model, the magnetic field of the circumbinary disk model also has  a poloidal component that influences the circularly polarized fraction if the disk has no edge-on inclination (see Fig.~\ref{fig:magnetic_field_cuts_mhd}, lower right). To investigate this, we performed simulations of the circumbinary disk model with an inclination from $0^\circ$ up to $90^\circ$ in $5^\circ$ steps. In contrast to Fig.~\ref{fig:cn_profile_mhd_simulation} (lower), we estimated the LOS magnetic field strength for both sides of the spectral line profiles without reducing the fitting range.

\begin{figure}
    \centering
    \resizebox{\scaling\hsize}{!}{\includegraphics{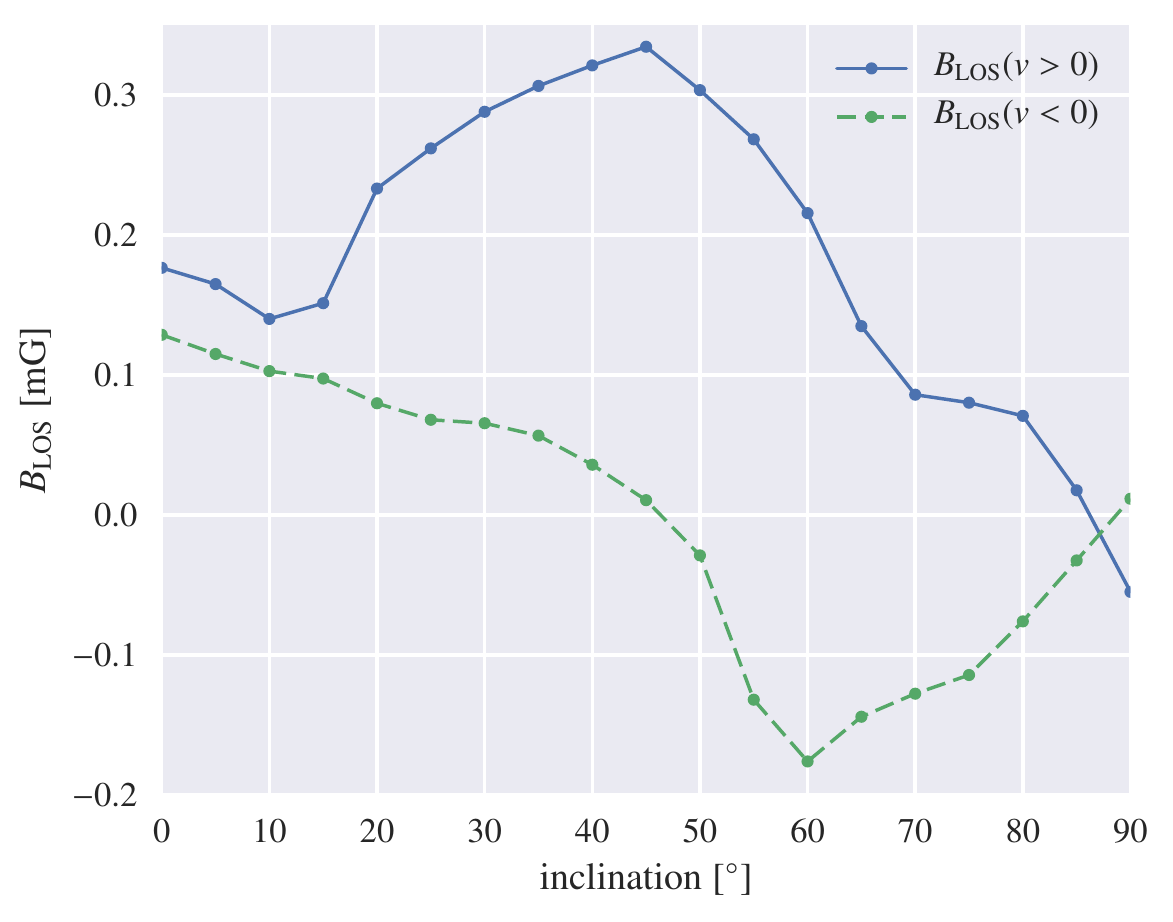}}
    \caption{Derived LOS magnetic field strength calculated from one half of the simulated $F_\text{I}$ and $F_\text{V}$ spectra of the circumbinary disk model based on a MHD simulation with different inclinations ($\SI{113.144}{GHz}$ CN line emission). For the solid/dashed line, only the part of the $F_\text{I}$ and $F_\text{V}$ spectra with positive/negative velocities is considered.}
    \label{fig:cn_profile_mhd_inc_study}
\end{figure}

As expected, both sides of the spectrum lead to a similar derived LOS magnetic field strength if the disk has a face-on inclination (see Fig.~\ref{fig:cn_profile_mhd_inc_study}). With increasing inclination, the side related to the positive LOS magnetic field strength of the toroidal field maintains the positive LOS magnetic field strength. In contrast, the side related to the negative LOS magnetic field strength acts against the poloidal magnetic field component and decreases the LOS magnetic field strength. As a result, differences between the derived LOS magnetic field strength of both sides of the spectral line profiles can also be explained with a poloidal component of the magnetic field. If the LOS magnetic field strength as a function of inclination is not known, it is almost impossible to distinguish between a poloidal component or asymmetries of the toroidal magnetic field. A solution for this ambiguity would be Zeeman observations of similar circumstellar disks with different inclinations. However, for inclinations ${\gtrsim}60^\circ$, the velocity shift is large enough to separate the circularly polarized flux of regions I and II in Fig.~\ref{fig:cn_profile_mhd_simulation}. In this case, the fitting process has a significant influence on the derived LOS magnetic field strength. As a consequence of fitting each side of the spectral line profiles in total, the derived LOS magnetic field strength converges towards zero for inclinations ${\gtrsim}60^\circ$ (see Fig.~\ref{fig:cn_profile_mhd_inc_study}). This result also affects  circumstellar disks around a single star since circumstellar disks in general are expected to possess a poloidal component of their magnetic field.

Another difference between the analytical circumstellar disk model with a single star and the circumbinary model based on a MHD simulation is the degree of circular polarization. It is up to two orders of magnitude smaller for the circumbinary disk model. This is partially caused by the inner region of the circumbinary disk ($R\lesssim\SI{20}{AU}$), which is thinned out by the gravitational forces of the two stars \citep[e.g.,][]{avramenko_observability_2017}. However, this effect only occurs if the separation between the two stellar components is on the order of ${\sim}\SI{10}{AU}$. A smaller separation would lead to a behavior that is similar to circumstellar disks around a single star. If the separation is significantly larger, the dust and gas around the binary can no longer be described with a single circumbinary disk. In circumstellar disks with a single star, the inner region has the highest density and usually the highest magnetic field strength \citep[see Fig.~\ref{fig:magnetic_field_cuts} in Sect. \ref{magnetic_field};][]{flock_3d_2017}. Therefore, as mentioned in Sect. \ref{result_mag_field}, circumbinary disks lack a significant fraction of circularly polarized emission if the inner high magnetic field region is missing. In addition, the different directions of circulation of the toroidal magnetic field decrease the circularly polarized emission from regions with similar velocities relative to the observer if not spatially resolved. Furthermore, the estimation of the LOS magnetic field strength from circumstellar disks with a single star is dominated by the inner region characterized by high magnetic field strength. In contrast, a Zeeman observation of a circumbinary disk is able to provide information about the magnetic field farther out in the disk if the requirements for such an observation are fulfilled. However, this requires that a significant part of its innermost region be depleted by the gravitational forces of both stellar components.

\section{Conclusions}\label{conclusions}
We studied the requirements to observe the Zeeman splitting of the $\SI{113}{GHz}$ CN lines in circumstellar disks and their dependence on selected disk parameters. Such observations can be used to resolve degeneracies and obtain complementary information about the magnetic field if combined with polarimetric continuum observations. However, this is only applicable, if the polarimetric continuum observations are mainly caused by elongated aligned dust grains. In particular, we evaluated whether Zeeman observations of circumstellar disks should be performed spatially resolved or unresolved.

With spatially resolved Zeeman observations, the strength and structure of the magnetic field in the LOS direction can be reconstructed for most regions in circumstellar disks. Only the region close to the star cannot be traced. However, the required sensitivity ($\Delta E\sim\SI{1}{\micro Jy}$, if $\Delta x\sim\SI{10}{AU}$) is far beyond the capabilities of  recent and upcoming instruments/observatories.

For spatially unresolved Zeeman observations, we obtained different results. Under the assumption of Keplerian rotation and a toroidal structure of the magnetic field, two independent average magnetic field strengths can be obtained from the circularly polarized fraction. These average values are sensitive to changes in the strength and structure of the magnetic field inside the disk. In addition, they can be used to detect major asymmetries in the toroidal structure of the magnetic field. However, these asymmetries can only be distinguished from a poloidal field component if the disk has an edge-on inclination or the magnetic field strength as a function of inclination is known to some extent. Such an observation would require a sensitivity that is  close to what recent and upcoming instruments/observatories are capable of ($\Delta E\sim\SI{0.1}{mJy}$). Nevertheless, multiple  circumstellar disks have an impact on this required sensitivity:
\begin{itemize}
    \item The magnetic field strength affects the required sensitivity in direct proportion. Therefore, circumstellar disks with relatively high magnetic field strengths in their surroundings, compared to other disks, should be preferred for Zeeman observations.
    \item The disk density and $CN/H$ abundance have a weak impact on the required sensitivity since circumstellar disks are usually partially optically thick. As a result, circumstellar disks with large amounts of CN molecules should be preferred, but not over disks that have a stronger magnetic field instead.
    \item A lower inclination of the disk reduces the required sensitivity by increasing the emitting cross section of the disk. However, deriving the LOS magnetic field strength from the circularly polarized fraction becomes more difficult with lower inclinations. Therefore, circumstellar disks with an inclination around $60^\circ$ should be preferred for Zeeman observations, while disks with inclinations below ${\sim}15^\circ$ should not be considered.
    \item The central star mass and therefore the Keplerian rotation has almost no influence on the required sensitivity. However, it changes the minimum inclination that is needed to obtain the average LOS magnetic field strength from the circularly polarized fraction.
\end{itemize}
We found that our magnetic field analysis is also applicable on a more complex and realistic circumbinary disk model that is based on a MHD simulation. Nevertheless, our simulations were based on a certain type of circumbinary disks. Therefore, the following results are only applicable to circumbinary disks with a separation of their stellar components on the order of ${\sim}\SI{10}{AU}$. Zeeman observations of circumbinary disks are more challenging in terms of sensitivity than observations of circumstellar disks with a single star. However, information about the magnetic field farther out in the disk  ($R>\SI{20}{AU}$) can be obtained with Zeeman observations of circumbinary disks. We also found that significant variations of the magnetic field strength close to the star (e.g., dead zones) cannot be detected with spatially unresolved Zeeman observations of the $\SI{113}{GHz}$ CN lines. 

We obtained most of our results from the simulated emission of the $\SI{113.144}{GHz}$ CN line. It is important to  note that not every spectral line at ${\sim}\SI{113}{GHz}$ has the same relative intensity and Zeeman shift. Therefore, the required sensitivity  to successfully perform a Zeeman observation of other CN lines at ${\sim}\SI{113}{GHz}$ might be higher. 

In summary, Zeeman observations of the $\SI{113}{GHz}$ CN lines have a great potential to increase the knowledge about the structure and strength of magnetic fields in circumstellar disks. Even if the requirements are extremely high for recent instruments/observatories, these observations should be achievable in the future. One goal of this study was to optimize Zeeman observations of circumstellar disks by determining how substantial disk parameters influence the requirements of such observations. For instance, the requirements would be significantly reduced for an observation of a large circumstellar disk with abundant CN, an inclination of about $60^\circ$, and magnetic field strengths of several $\SI{100}{\micro G}$ in its surrounding ($CN/H>\SI{e-7}{}$, $R_\text{out}>\SI{300}{AU}$).

 \begin{acknowledgements}
       Part of this work was supported by the German
       Deut\-sche For\-schungs\-ge\-mein\-schaft, DFG\/ project number WO 857/12-1 on circumbinary disks.
 \end{acknowledgements}

\nocite{hunter_matplotlib:_2007}
\bibliographystyle{aa}
\bibliography{my_library,custom_bibtex}
\end{document}